\documentclass[submission, Phys]{SciPost}
\usepackage{hyperref,color}
\usepackage{amsmath,amssymb,bm}
\usepackage{hyperref}
\usepackage{graphicx}
\graphicspath{{figureN/}}
\usepackage[caption=false]{subfig}
\usepackage{float}
\usepackage{subfiles}
\usepackage{commath}
\usepackage[normalem]{ulem}
\usepackage{cancel}
\usepackage{enumitem}
\usepackage{indentfirst}

\begin{document}
\begin{center}
    \Large{ \textbf{Time-reversal-broken  Weyl semimetal in the Hofstadter regime}}
\end{center}

\begin{center}
Faruk Abdulla\textsuperscript{1*},
Ankur Das\textsuperscript{2$\dagger$},
Sumathi Rao\textsuperscript{1,3},
Ganpathy Murthy\textsuperscript{4}
\end{center}

\begin{center}
{\bf 1} Harish-Chandra Research Institute, HBNI, Chhatnag
Road, Jhunsi, Allahabad, India
\\
{\bf 2} Department of Condensed Matter Physics, Weizmann
Institute of Science, Israel
\\
{\bf 3} International Centre for Theoretical Sciences, Tata Institute
of Fundamental Research, Bengaluru, India
\\
{\bf 4} Department of Physics and Astronomy, University of 
Kentucky, Lexington, KY, USA
\\
*farukabdulla@hri.res.in,
$\dagger$ankur.das@weizmann.ac.il
\end{center}

 \newcommand{\FA}[1]{{\color{orange}{ FA: #1}}}
 \newcommand{\FAEDITED}[2]{{\color{orange} \sout{ #1 }}{ \textcolor{orange}{#2}}}

\newcommand{\AD}[1]{{\color{blue}{ AD: #1}}}
\newcommand{\ADEDITED}[2]{{\color{blue} \sout{ #1 }}{ \textcolor{blue}{#2}}}

\newcommand{\SR}[1]{{\color{red}{ SR: #1}}}
\newcommand{\SREDITED}[2]{{\Rout{#1}}{ \textcolor{red}{#2}}}

\newcommand{\GM}[1]{{\color{red}{\textbf{GM:#1}}}}
\newcommand{\GMEDITED}[2]{{\color{red}\sout{#1}}{\textcolor{red}{#2}}}

\section*{Abstract}

We study the phase diagram for a lattice model of a time-reversal-broken
three-dimensional Weyl semimetal (WSM) in an orbital magnetic field $B$ with
a flux of $p/q$ per unit cell ($0\le p \le q-1$), with minimal crystalline
symmetry. We find several interesting phases: (i) WSM phases with $2q$, $4q$,
$6q$, and $8q$ Weyl nodes and corresponding surface Fermi arcs, (ii) a layered
Chern insulating (LCI) phase, gapped in the bulk, but with
gapless surface states, (iii) a phase in which some bulk bands are gapless
with Weyl nodes, coexisting with others that are gapped but topologically 
nontrivial, adiabatically connected to an LCI phase, (iv) a new gapped trivially
insulating phase (I$'$) with (non-topological) counter-propagating surface states,
which could be gapped out in the absence of crystal symmetries.
Importantly, we are able to  obtain the phase boundaries analytically for all
$p,q$. Analyzing the gaps for $p=1$ and very large $q$ enables us to smoothly take the 
zero-field limit, even though the phase diagrams look ostensibly very different
for $q=1, B=0$, and $q\to\infty, B\to 0$.


\textcolor{magenta}{\section{Introduction}}
\label{introduction}

Weyl semimetals (WSM) \cite{Shuichi_Murakami_2007,Savrasov_etal_2011,
Vishwanath1_etal_2012,Ran_etal_2011,Balents_etal_2011,Fang_etal_2011,
Hasan1_etal_2015,Hasan2_etal_2015, Ding1_etal_2015,Ding2_etal_2015,
Soljacic_etal_2015} are examples of topological quantum
matter which are not fully insulating in the bulk but have an even
number of points in the Brillouin zone called Weyl nodes where the
conduction and valence bands touch each other. Each Weyl node has a
topological charge and acts as a source or sink of the Berry phase in
momentum space. The gapless surface states of the WSM consist of Fermi
arcs in the surface Brillouin zone (SBZ) which join the projections of
the Weyl nodes onto the SBZ.

It is well-known
\cite{Burkov_Hook_2011,Spivak_etal_2013, Analytics_etal_2016,Li_Roy_DasSharma_2016} 
that in the presence of parallel external electric
and magnetic fields, the density of electrons at individual
nodes is not conserved and transport in the WSM is anomalous due to the chiral
anomaly. This causes the electrons to be pumped from
one Weyl node to another with opposite topological charge until this
process is balanced by  internode scattering. The surface Fermi
arcs also show interesting quantum oscillations
\cite{Potter_etal_2014,Zhang_etal_2016,Wang_Zhou_2017,Zhang_2019} in an applied
magnetic field. Semiclassically, the Lorentz force causes the electrons
to move along the Fermi arc, tunnel through the bulk at the Weyl node
and then complete the circuit via the Fermi arc on the other surface and
tunneling back through the bulk.

Beyond the semiclassical limit, the effect of orbital magnetic fields
$B$ on the WSM has been well-studied \cite{Burkov_Hook_2011,Li_Roy_DasSharma_2016}
in the continuum limit (magnetic length $\ell = \sqrt{\hbar/eB}$ is much larger
than the lattice spacing $a$), where the $B$ field is weak enough that attention
can be restricted to states very close to the Weyl nodes. Coming to
strong fields, Roy and co-workers \cite{Grushin_etal_2016} have characterized
the Hofstadter-Weyl butterfly in a simple two-band model with two Weyl
nodes, where the magnetic field is applied parallel to the separation
between the Weyl nodes. There has also been some work \cite{Gao_Li_Zhang_2018}
on obtaining the energy spectra for the
Hofstader  Hamiltonians for Weyl and double-Weyl semimetals, where the
systems are shown to exhibit 3D quantum Hall effect
\cite{Bernevig_etal_2007,Yang_Lu_Ran_2011,Wang_Zhou_2017} for appropriate hopping
parameters and rational fluxes.

We go beyond previous work in two ways. First,  we consider a
two-band model with a mimimal  crystalline symmetries,
- i.e., with fully anisotropic hoppings. Here, even without an external magnetic
field, we find that  there are several new phases  of WSMs with 2, 4, 6, and 8
Weyl nodes. Earlier studies\cite{Yang_Lu_Ran_2011,Chen_etal_2015,Hassan_Taylor_2016,Roy_etal_2018}
had imposed crystalline symmetries and had uncovered the WSM phases with 2 and 4 nodes
with the LCI phase (also sometimes called a 3D Quantum Hall state 
\cite{Chen_etal_2015} or a  3D Chern Insulator \cite{Hassan_Taylor_2016}),
but the phases with 6 and 8 nodes had not been seen earlier and
requires unconstrained anisotropic hoppings.  Second, unlike earlier
studies of the effect of magnetic fields on this model, we consider the
case when the external $B$ field is perpendicular to the separation
between the nodes.  Constraining one of the hopping parameters, we are 
able obtain the phase boundaries analytically at arbitrary values of $p,q$.
Last, but not least, we smoothly connect our results for small but
commensurate fields to the semiclassical limit.

\begin{figure}[ht]
\centering
\includegraphics[width=1\linewidth]{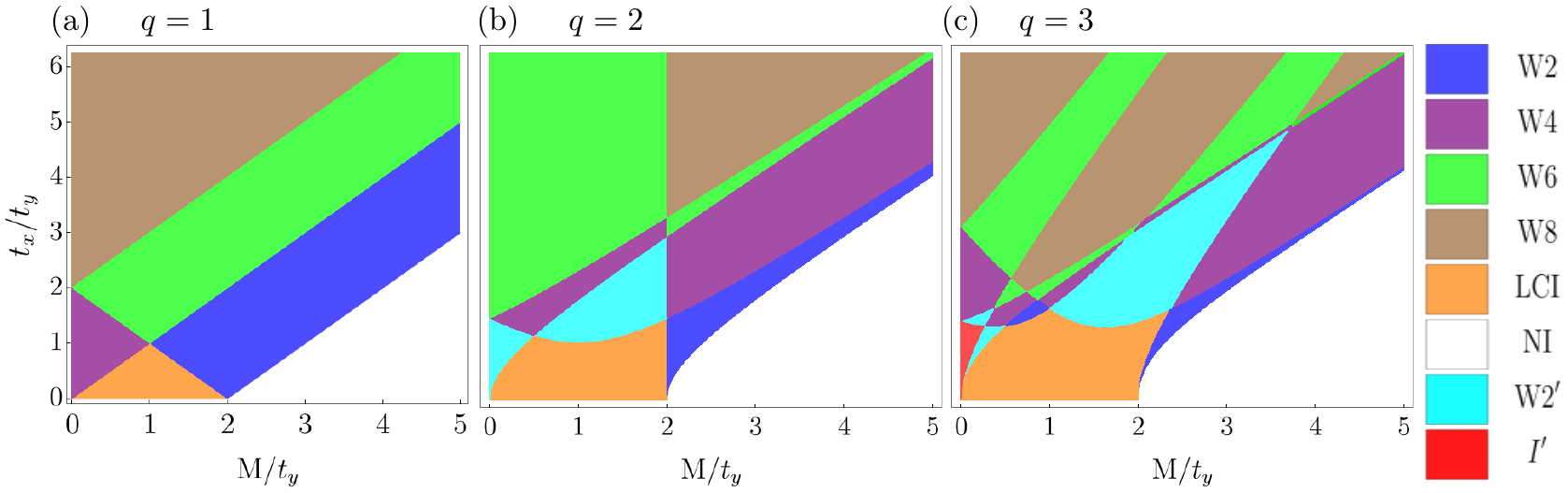}
\caption{The phase diagram for (a) zero field ($q=1$) and (b)-(c) for
finite commensurate fields ($q=2, 3$), with $t^{(1)}_z=t^{(1)}_y =t_y
$. (a) The gapless phases consisting of WSM phases with 1, 2, 3, and
4 pairs of Weyl nodes are shown in blue, purple, green, and
brown. The gapped phase (in orange) is a layered Chern insulator
(LCI), where the 2D Chern layers are coupled via $t_x$. The white
region at the right bottom corner represents a trivial insulator. 
In (b)-(c) the WSM phases  have $2q$, $4q$, $6q$, and $8q$ Weyl nodes in the
bulk, labeled W2, W4, W6, and W8 respectively. In non-zero magnetic fields,
two  additional phases appear which are labeled  W2$'$ and $I'$. The W2$'$ 
phase (aqua) is gapless and has $2q$ Weyl nodes, but coexists with an LCI,
and hence possesses  gapless surface modes in addition to the usual Fermi
arcs. The $I'$ phase is a gapped, topologically trivial, insulating phase
but has counterpropagating zero-energy surface states. We also note that
the gapped LCI and $I'$ phases take up  regions from the WSM phases and
expand with increasing $q$. We emphasize that the phases of the same color
in (a), (b), and (c) are not identical. The number of  Weyl nodes in 
corresponding phases  is related by the factor $q$.}
\label{fig:phase}
\end{figure}

Our central result is the set of phase diagrams in Fig. \ref{fig:phase} 
for each $q$ (where $p/q$ quanta of flux go through each unit cell).
The phase diagrams are universal in two distinct ways: (i) For each
value of $q$, the set of phases and the regions they occupy in parameter
space are independent of $p$. (ii) For all $q\geq1$, phases with 
$2q$, $4q$, $6q$, and $8q$ Weyl nodes appear in the phase diagram, as well as the
trivial insulator and the LCI phase. For all $q\geq2$, a phase in which gapless Fermi
arcs coexist with gapless chiral states spanning the surface BZ (characteristic of the
LCI) always appears. For all $q\geq3$ a topologically trivial insulator with gapless
surface modes on certain surfaces (protected by translation symmetry) always appears. 

In a particular gauge, the $q$-fold increase in the number of Weyl nodes is
essentially a consequence of the $q$-fold increase in the periodicity along the $y$ axis.
This $q$-fold increase in periodicity also leads to an increase in the number of phase 
boundaries and consequently an increase in the number of times each phase appears in the
phase diagram. 
Transitions between phases with differing numbers of Weyl
nodes occur via the creation/annihilation of pairs of Weyl nodes of
opposite topological charge. The bulk dispersion at the location of
the phase transition is quadratic in one direction but linear in the other
two. The corresponding surface spectrum of the Fermi arcs is also quadratic in
one direction but linear in the other at the phase transitions.

While the phase diagram is independent of $p$, the spectrum strongly depends on $p$.
For a certain restricted, but still nontrivial, set of parameters,
we can solve for the phase boundaries analytically, which enables us to go to very 
large values of $q$,  where the phase diagram in most of the parameter space
approaches a limit as $q\to\infty$. The exception is a complicated patchwork of many
phases occurring in  a narrow band of parameters, whose width vanishes as
$q\to\infty$ (Fig. \ref{fig:Largeq}). At first glance, it is puzzling that the
phase diagram as $p=1,q\to\infty$ (Fig. \ref{fig:Largeq}) which is the semiclassical
limit $B\to0$, looks very different from that at 
at $B=0$ (Fig. \ref{fig:phase}a). However, a detailed examination of the
asymptotic behavior of the  gap near zero energy enables us to make the correspondence
between the $p=1,q \to\infty$ and the $B=0$ phase diagrams (studied in Figs. \ref{fig:Largeq}-\ref{fig:bandxyz}). 

Many of the features that we find in our analysis can be translated to experimental
detection of the new phases via their topological responses. For instance, phases with 
co-existence of the LCI  and a WSM can lead to both of them contributing to the Hall
conductance (Figs. \ref{fig:spectrum}-\ref{fig:cartoon}). The Hall conductance can also
be tuned by changing parameters to go between phases.

The plan of the paper is as follows. In Sec. 2, we introduce the
fully anisotropic two-band lattice model of a WSM with broken
time-reversal symmetry and obtain its phase diagram at zero flux. In
Sec. 3 we obtain the phase boundaries analytically for $p/q$ flux
quanta piercing each unit cell in a direction perpendicular to the
Weyl-node separation (the $x$ direction). In Sec. 4 we study the
bulk and the surface spectra, paying particular attention
to gapless surface states. In Sec. 5 we discuss the weak-field limit
in detail and make contact with the continuum description. We
conclude in Sec. 6 with a summary, potential caveats in our results,
the effects of disorder, and possible directions for the future. A
number of  straightforward mathematical details are relegated to the
appendices.


\textcolor{magenta}{\section{The lattice model and its phase diagram at zero flux} 
\label{zerofluxmodel}}

We begin with a time-reversal-broken two-band lattice model
\cite{Yang_Lu_Ran_2011} of Weyl semi-metal defined on a cubic lattice
given by the following Hamiltonian

\begin{align}
\begin{aligned}\label{eq:realspaceH}
H = \sum_{{\bf n}, j} c^{\dagger}({\bf n}) 2M \sigma_x c({\bf n}) 
-\left(c^{\dagger}({\bf n }+ a\hat{e}_{j}) ~ T_{j} ~ c({\bf n}) + H.c.
\right) 
\end{aligned}
\end{align}
where ${\bf n} = a(n_x, n_y, n_z)$, $n_i$ being integers, denote the lattice sites,
$\hat{e}_j$ is the unit vector along $j^{th}$  direction, the $\sigma_i$'s
are the Pauli matrices representing the (pseudo)spin and ${\bf c}
({\bf n})=(c_\uparrow({\bf n}), c_\downarrow({\bf n}))^T$ are the two-component fermions. 
Microscopically, the pseudospin label arises from spin-orbit coupling,
leading to eigenstates which are  linear combinations of spin and
orbital eigenstates. The hopping matrices $T_j$,  $j = (x, y, z)$, are given by:
$T_x = t_x \sigma_x$,  $T_y = t^{(1)}_y \sigma_x +i t^{(2)}_y\sigma_y$  and 
$T_z = t^{(1)}_z \sigma_x + i t^{(2)}_z\sigma_z$. The lattice constant $a$ is
set to be unity for the rest of the paper. This model has been studied earlier for 
isotropic hoppings $t_x=t^{(1)}_y=t^{(1)}_z=t_1$, $t_y^{(2)}=t_z^{(2)} = t_2$
and with $t_1=t_2$ and is known\cite{Yang_Lu_Ran_2011,Chen_etal_2015,Hassan_Taylor_2016,Roy_etal_2018}
to have a Weyl semimetal (WSM) phase for $|M/t_1|<3$. In this paper we  work 
with fully anisotropic hoppings, where the model has  minimal rotational symmetry
- a single two-fold rotation about the $x$-axis (the full symmetry analysis is
carried out in the Appendix \ref{appendix:symmetryanalysis}).

To set the stage for our nonzero flux results, we first analyse the model 
at zero flux in the full parameter space. We find that the topological phase
diagram, in general, depends only on three of the hoppings ($t_x,t^{(1)}_y,
t^{(1)}_z$) and the onsite mass parameter $M$, which are all real. 
The Bloch Hamiltonian in ${\bf k}$-space is
\begin{align}\label{eq:blochH1}
\begin{aligned}
h({\bf k}) = ~ & 2\left(M - t_x \cos{k_x} - t^{(1)}_y \cos{k_y} - t^{(1)}_z
\cos{k_z}\right)\sigma_x  \\
& + 2t^{(2)}_y \sin{k_y}\sigma_y + 2t^{(2)}_z \sin{k_z}\sigma_z.
\end{aligned} 
\end{align}
Since the second and third terms vanish at $k_y=0, \pi$ and $k_z=0, \pi$
respectively, the gapless points in the energy spectrum along with $k_x$
directions  are given by the zeroes of the first term
\begin{align}\label{eq:B0gapless}
\cos{k_x} = \left(M - (-1)^{\mu}~ t_y^{(1)} - (-1)^{\nu} ~ t_z^{(1)} \right)/t_x .
\end{align}
Note that here both $\mu$ and $\nu$ take values 0 and 1, depending on
the $k_y$ and $k_z$ values being $0$ and $\pi$ respectively. The
above equation represents a set of four equations whose solutions are
the Weyl nodes of the model, each equation potentially producing a pair
of Weyl nodes. Since both $\mu$ and $\nu$ take two values each, and each
equation can have two values of $k_x$ where the spectrum can be
gapless, there could be upto eight distinct Weyl nodes. Of course, not
all parameter values support eight solutions. We label the WSM phases
with 1, 2, 3 and 4 pairs of Weyl nodes as W2, W4, W6, and W8 in
Fig. \ref{fig:phase}(a) respectively.

We now show how to obtain the phase boundaries in the $(M, t_x)$ parameter space 
from Eq. \ref{eq:B0gapless}. The knowledge we gain from here will be useful
when we will study the model in a magnetic field in the following sections.
The left hand side of Eq. \ref{eq:B0gapless} is 
$\cos{k_x}$ which lies in the range  $[-1,1]$. Equation \ref{eq:B0gapless} has
a solution only in the range of parameters for which the right hand side also
lies in the same range for a given $\mu, \nu$. Recall that different $\mu,\nu$
correspond to different values of $k_y,k_z$ where the spectrum is gapless. 
Thus, Weyl  nodes at different $\mu,\nu$
cannot annihilate each other. A phase transition corresponds to a change in the
number of Weyl nodes at a specific $\mu,\nu$, which can only occur when two Weyl
nodes are either created or destroyed. Eq. \ref{eq:B0gapless} shows that if $k_{x0}$
is a solution, so is $-k_{x0}$. The two solutions can be created/annihilated only
when they coincide in the BZ, which means at $k_{x0}=0,\pi$.  Therefore,  at a
putative phase transition, the right hand side of Eq. \ref{eq:B0gapless} should
be at the boundary of its allowed range, namely $\pm1=(-1)^\delta$. Thus the
condition for the phase transition is,

\begin{align}\label{eq:critical0}
& M - (-1)^{\delta}~ t_x  =  (-1)^{\mu} t^{(1)}_y + (-1)^{\nu} t^{(1)}_z .
\end{align}
The above equation represents a set of eight equations because
each of the exponent $\delta$, $\mu$ and $\nu$ take the two values 0 and 1.
Clearly, the equations are symmetric under $M \to -M$ and $t_x \to -t_x$,
because this does not lead any new conditions - it merely shuffles the
set of eight equations. Therefore, we can restrict ourselves to $M, t_x
\geq 0$.

In addition to WSM phases and a trivial insulator, we find a layered
Chern insulator (LCI) with a bulk gap. The LCI phase can be imagined
as a stack (along the $x$-direction) of 2D Chern insulator layers in the
$yz$ plane which are tunnel-coupled via $t_x$. When $t_x=0$, we have a
set of disconnected 2D Chern insulator layers for $M<2$ (for $M>2$, it
is a trivial insulator, see Fig. \ref{fig:phase}a), with a Chern number
$C=1$ per layer. For $t_x
\neq 0$, this simple picture no longer holds, but one can compute the
Chern number $C(k_x)$ for each value of $k_x$ by integrating the Berry
curvature over the two dimensional $k_y$-$k_z$ BZ. We obtain
$C(k_x)=1$ for each $k_x$ showing that the LCI is adiabatically
connected to the $t_x=0$ limit. The LCI phase 
has the expected surface state, which is the collection of quantum
Hall-like chiral edge states of the 2D Chern insulators, spanning the
entire Brillouin zone along $k_x$ direction. Here we note  that increasing 
tunnel-coupling $t_x$ runs a phase transition from LCI to WSM \cite{Burkov_Balents_2011}
(also clear from  the phase diagram Fig. \ref{fig:phase}a ).

The Fermi arc surface states in all the WSM phases as well as in the
LCI phase are (pseudo)spin-polarized along the $z$-direction. The surface states which are
localized on the $xy$ crystal surface have the low energy
dispersion $E(k_x, k_y) \propto k_y$, and the surface states which are
localized on the $xz$ crystal surface have the low energy
dispersion $E(k_x, k_z) \propto k_z$.  For completeness, we have shown
the spectra for some of the phases in Appendix \ref{appendix:sim}. We
have also shown plots of a few surface states and numerically 
confirmed their pseudospin polarizations and the directions of their
velocities.

\textcolor{magenta}{\section{The phase diagrams for \texorpdfstring{$p/q$}{p/q}
flux quanta per unit cell\label{sec:phasediag}}}
\label{phasedias}

\vspace{0.4cm} 

\subsection{Hamiltonian in Hofstadter regime}

Now we are ready to examine how the phase diagram at zero orbital flux
(Fig. \ref{fig:phase}a) gets modified in the presence of an orbital
magnetic field. Note that we consider the orbital effect only. The
reason is the following: The orbital coupling, since it couples to the
charge degree of freedom, is universal. On the other hand, since the
pseudospin label is a ${\bf k}$-dependent linear combination of spin
and orbital labels, which does depend sensitively on the microscopic
material parameters, the associated Zeeman coupling is not universal.
In any case, the most general form of the Zeeman
coupling in our Hilbert space is
\begin{align}\label{eq:Zeeman}
H_Z=\gamma(\mathbf{k}) B{\hat n}(\mathbf{k}) \cdot {\boldsymbol\sigma} 
\end{align}
Provided $\gamma(\mathbf{k})$ is not too large, the effect of the Zeeman coupling
will be to shift the locations of the Weyl nodes (if present), and
shift the phase boundaries. Since the Zeeman coupling is not expected
to introduce anything qualitatively different, we will ignore it in the
following.

We consider an orbital flux along the $z$-direction perpendicular to
the $xy$ plane. The hopping terms in the Hamiltonian pick
up a nontrivial phase factor under Peierls substitution\cite{Peierls_1933}.
We will work in the Landau gauge ${\bf A} = (-y, 0, 0)B$, where only the
hopping in the $x$-direction picks up a nontrivial phase so that the
Hamiltonian  in a magnetic field is obtained from Eq. \ref{eq:realspaceH} by the 
replacement $T_x \to  T_x \exp(-i 2 \pi y\phi/\phi_0)$. We will restrict 
ourselves to the case where the flux $\phi$ (in units of the quantum flux
$\phi_0=h/e$) per unit cell is commensurate i.e. $\phi=Ba^2/\phi_0 = p/q $,
where $p$ and $q$ are relatively prime, so that translation symmetry along
the $y$-direction is  restored with a larger unit cell \cite{Hofstadter_1976}.
To diagonalize the Hamiltonian, we define a magnetic unit cell which is $q$
times the original unit cell, extended in the $y$-direction. Upon Fourier
transformation with respect to the Bravais lattice sites of the magnetic unit
cell, the following Bloch Hamiltonian is obtained 
\begin{align}\label{eq:blochH2}
h_{\phi}({\bf k})  = \sum_{\alpha=0}^{q-1} & c_\alpha^{\dagger}({\bf
k})\left[f_{1}^{\alpha}({\bf k})\sigma_x + f_{3}^{\alpha}({\bf k})\sigma_z\right]
c_\alpha^{}({\bf k}) \nonumber \\
& - \left(c_{[\alpha+1]}^{\dagger}({\bf k}) e^{i q k_y \delta_{(\alpha,q-1)}} ~ T_y ~
   c_\alpha^{}({\bf k}) + H.c.\right). 
\end{align}
We will refer to $h_{\phi}({\bf k})$  as the Hofstadter Hamiltonian, where   
$\alpha = 0, 1, ...,q-1$  are the sublattice indices in  the magnetic unit cell 
and $\mathbf{k}$ lies in the reduced (magnetic)  Brillouin zone,
$i.e.$, ${\bf k}$: $k_x \in \left(0, 2\pi\right)$,   $k_y \in \left(0, 2\pi/q
\right)$, $k_z \in \left(0, 2\pi\right)$. The square bracket notation in the above
equation implies that the values are taken modulo $q$ - $i.e.$, $[\alpha]= \alpha
\mod q$. The hopping matrix  $T_y = t^{(1)}_y\sigma_x + it^{(2)}_y \sigma_y$ has been
defined  earlier in the previous section. The functions $f^\alpha_1$ and $f^\alpha_3$ are
defined as
\begin{subequations}
\begin{align}
f^{\alpha}_1({\bf k}) =& 2 \left(M - t_x\cos{\left(k_x + \frac{2\pi p}{q}
\alpha\right)}
- t^{(1)}_z \cos{k_z}\right)\\ 
f^{\alpha}_3({\bf k}) =& 2 t^{(2)}_z \sin{k_z}.
\end{align}
\end{subequations}
Note that the spectrum of $h_{\phi}({\bf k})$, shown explicitly in
Appendix \ref{appendix:qSol}, is particle-hole symmetric. As shown in
Appendix \ref{appendix:qSol}, we can compute the entire phase diagram
analytically if we set  $|t^{(2)}_y|=|t^{(1)}_y|$. To avail
ourselves of the simplicity and computational advantages this gives us,
especially at large $q$, we will make
this choice $t^{(2)}_y=t^{(1)}_y \equiv t_y$ for the rest of the paper.

\subsection{Gapless points and phase boundaries}

To obtain the phases and phase boundaries, we need to identify the
zeroes of the Hofstader Hamiltonian $h_\phi(\mathbf{k})$ (its spectrum
is particle-hole  symmetric about the zero energy as shown in  Appendix \ref{appendix:symmetryanalysis}), which gives
us the band-touching points where the energy spectrum vanishes, which
in turn allows us to find the number of Weyl nodes. This can be done
by explicitly writing the Bloch Hamiltonian as a $2q\times 2q$ matrix
in the basis of sublattice and  pseudospin $\Psi =
\left(c_{0,\uparrow}, c_{1,\uparrow }, ..., c_{q-1,\uparrow},
c_{0,\downarrow}, c_{1,\downarrow }, ..., c_{q-1,\downarrow}
\right)$. The details of the calculation are shown in Appendix
\ref{appendix:qSol}. The energy gap can close only at $k_y=0$
and/or $k_y=\pi/q$, and only at $k_z=0$ and/or
$k_z= \pi$. For each of these possibilities, the $k_x$ values at which
the spectrum is gapless are given by
\begin{align}\label{eq:gapCondition}
\cos{qk_x} = (-1)^{p+q}\left[ - T_q(g) + (-1)^{\mu -q}~ 2^{q-1} ~
(t_x/t_y)^{-q} \right],
\end{align}
where $\mu$ takes values 0 and 1 which correspond to closing of the gap at
$k_y=0$ and $k_y=\pi/q$ respectively. Here $T_q(g)$ is the Chebyshev
polynomial  \cite{abramowitz_Stegun_Romer_1988} of the first kind of degree
$q$. Its argument $g = \left(-M + \left(-1\right)^{\nu}
t^{(1)}_z \right)/t_x$, where $\nu$ takes values 0 and 1, which 
correspond to closing of the gap at $k_z=0$ and $k_z=\pi$ respectively. 
So essentially, Eq. \ref{eq:gapCondition} is a set of four independent
equations.  Note also that Eq. \ref{eq:gapCondition} involves only the
parameters $g$ and $t_x/t_y$. Finally, we note that for a given $(g,
t_x/t_y)$, if $k_x= k_0$ is a solution of  Eq. \ref{eq:gapCondition},
then so is $k_x=-k_0$. Furthermore, since $\cos{q(\pm k_0 + 2\pi m/q)} =
\cos{qk_0}$, it is easy to see that $k_x=k_0 + 2\pi m/q$ where $m=0, 1, 2,
...,(q-1)$ are also solutions. Since, without loss of generality, $k_0$
can be restricted to $0\leq k_0  \leq 2\pi/q$, it is clear that there
are $2q$ values of $k_x$ in the magnetic BZ where the spectrum could be
gapless.  Since $\mu$ and $\nu$ take two values each, and each
equation can have $2q$ values of $k_x$ where the spectrum can be gapless,
there could be regions in the parameter space with up to $8q$ distinct
gapless values for $k_x$. These distinct gapless points in the magnetic BZ
are the Weyl nodes in the theory. 

We can now obtain the phase boundaries from Eq.
\ref{eq:gapCondition} by arguments identical to those given in the zero
field case in Sec. \ref{zerofluxmodel} above Eq. \ref{eq:critical0}. As 
the left hand side is now changed
to $\cos{q k_x}$ the number of zeros will be $2q$ for a given $\mu, \nu$ and
the phase boundary equation will be (after setting RHS to $\pm 1$
in Eq. \ref{eq:gapCondition}),
\begin{align}\label{eq:gap} 
- T_q(g) + (-1)^{\mu -q}~ 2^{q-1} ~ (t_x/t_y)^{-q} = (-1)^{p+q+\delta}
\end{align}
where $\delta$ takes two values $0$ and $1$, which correspond to setting 
the RHS of Eq. \ref{eq:gapCondition} equal to $1$ and $-1$ respectively. 
Analogous to Eq. \ref{eq:critical0} in the zero-field case, the above
compactly
written equation is a set of eight independent equations because each
of the exponents $\mu$, $\nu$ and $\delta$ take two values. Note
that the only appearance of the numerator $p$ (of the flux $p/q$) is
in the exponent on the RHS, and the value only depends on whether
$p$ is odd or even. This only shuffles between the set of equations
and does not lead to any new condition. So the phase boundaries are
completely independent of the value of $p$.  Interestingly, as we
mentioned earlier, there are only two parameters $g$ and $t_x/t_y$
which enter the Eq. \ref{eq:gap}. The hopping parameter
$t^{(1)}_z $ enters the equation via the value of $g = \left(-M +
\left(-1\right)^{\nu} t^{(1)}_z \right)/t_x$, which, given that a
solution exists, determines the values of $k_x$ at which there are
Weyl touchings. To simplify our analysis without sacrificing anything 
qualitatively new, we henceforth fix $t^{(1)}_z =t^{(1)}_y= t_y^{(2)}=
t_y = 1$ and $t_z^{(2)}=1$. With this 
simplification, our phase diagram is  essentially controlled only by two 
parameters $M/t_y$ and $t_x/t_y$.  The same Eqs.
\ref{eq:gapCondition}-\ref{eq:gap} are also applicable to the zero field case
and correctly reproduce (setting $q=1$ in Eq. \ref{eq:gap}) the phase boundaries
given by Eq. \ref{eq:critical0}. \\

Note that  a uniform  magnetic field in the  $y$-direction,  
would give  identical conditions (Eqs. \ref{eq:gapCondition} 
and \ref{eq:gap}) for the gapless spectrum and the phase boundary respectively, if we make 
the replacements   $t_z^{(1)} \leftrightarrow t_y^{(1)}$, $t_z^{(2)}
\to -t_y^{(2)}$ and $t_y^{(2)} \to t_z^{(2)}$. So {all}  our results including the phase diagrams
described in the following  sections  apply to this case as well.


\begin{figure}[ht]
\centering
\includegraphics[width=0.8\columnwidth]{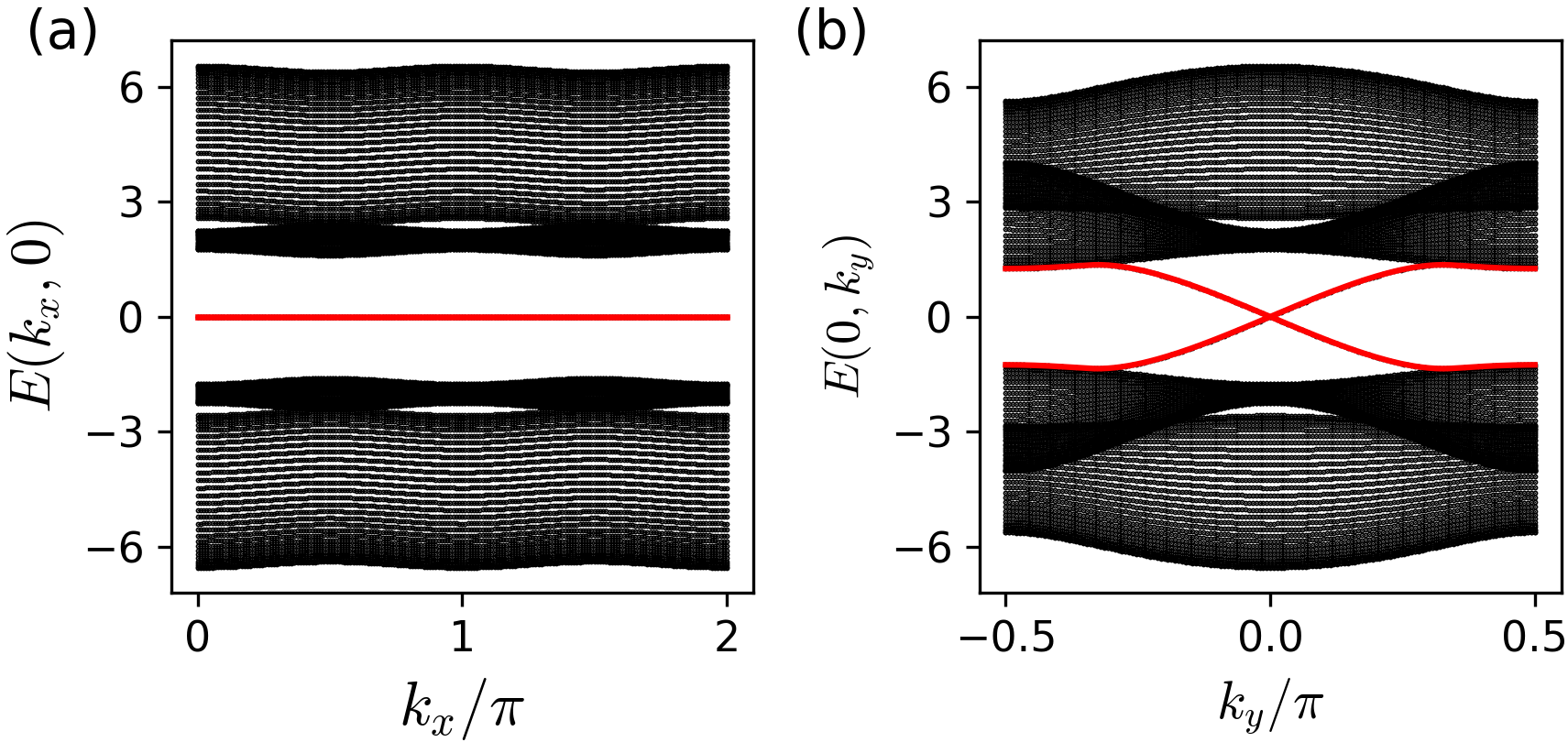}
\caption{The spectrum of a finite slab in the $z$-direction in the LCI
  phase, when $q=2$. The parameters are $M=1.2t_y$, $t_x=0.4t_y$.  (a)
  The surface states (highlighted in red) are dispersionless in $k_x$
  and span the BZ. (b) The surface states (highlighted in red)
  disperse in $k_y$, with linear dispersion near zero energy, with
  two gapless modes of opposite slopes belonging to opposite open
  surfaces of the slab. The surface state living on the $xz$ crystal surface,
  has  a linear dispersion in $k_z$ near zero energy (not shown here).}
\label{fig:LciSband} 
\end{figure}

\subsection{ Phase diagrams for a strong field, small q}

\begin{figure}[!ht]
\centering
\includegraphics[width=0.95\columnwidth]{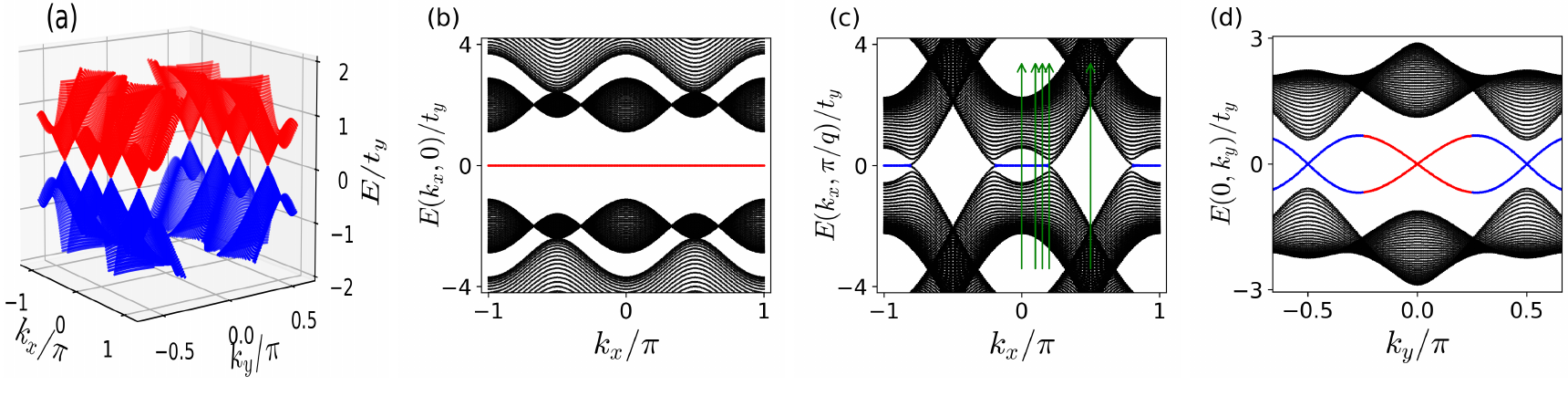}\\
\includegraphics[width=0.95\columnwidth]{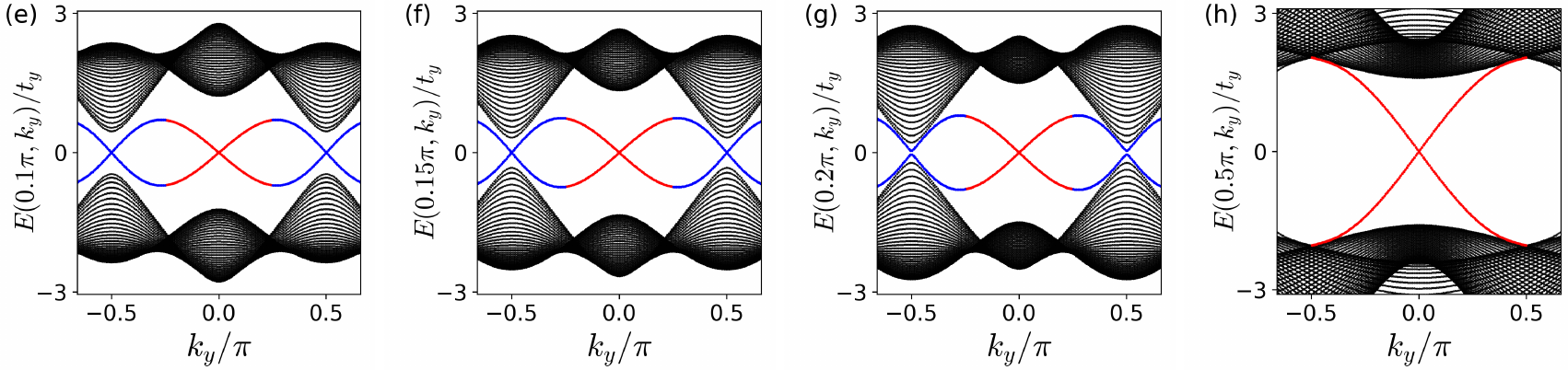}
\caption{Bulk and surface spectrum in the W2$'$
  phase for $q=2$, for $M=1.2t_y$ and $t_x=1.3t_y$. (a) The bulk
  bands around zero energy show $2q=4$ Weyl nodes which live at the
  boundary of the magnetic BZ ($k_y=\pm \pi/q$, $k_z=0$). In (b)-(h) the spectrum
  along $k_x$ and $k_y$ directions is shown for the WSM slab (finite in the
  $z$-direction with $L_z=40$). (b)Spectrum along $k_x$ for a fixed $k_y=0$ shows 
  the  surface states (in red) which are not connected to the bulk Weyl node projections
  and span the magnetic BZ. (c) Spectrum along $k_x$ for a fixed $k_y=\pi/q$ shows 
  the Fermi arc surface states (in blue) connecting  the bulk Weyl node projections.
  (d)-(h) Spectrum along $k_y$ for a series of values of $k_x$ (indicated in (c) by vertical 
  green arrow) shows the existence of two types of surface states. The surface states around
  $k_y=\pm \pi/q$ correspond to the Fermi arc surface states which are connected  to the
  bulk Weyl node projections, whereas those around $k_y=0$ are the additional surface states 
  unconnected to the bulk. The red and blue shadings  are meant  for visualizations of the
  surface states  around $k_y=0$ and $k_y=0.5\pi$ respectively. As the  Weyl point projection is 
  approached, the associated Fermi arc surface states start to mix with the bulk states (see figure (g))
  and the decay length will diverge when we hit the Weyl node. In (h) we have only one type of
  surface states. The decay length of the surface states (shown in Figs. \ref{fig:spectrum}e-h) 
  for a series of  $k_y$ values in between $k_y=0$ and $0.5\pi$ is plotted 
  in Fig. \ref{fig:surface}. }
\label{fig:spectrum}
\end{figure}

In the earlier section, we obtained an analytical expression for the
phase boundaries at any arbitrary commensurate magnetic flux
penetrating a primordial unit cell $\phi=Ba^2/\phi_0 = p/q $, where
$p$ and $q$ are relatively prime. As explained earlier, the phase
boundaries are independent of $p$, and the phase diagrams are shown
in Fig. \ref{fig:phase} for $q=1,2,3$.  Note that $q=1$ implies that an integer
quantum of flux pierces the primordial unit cell. Since there are no
nontrivial phase factors in the hopping terms, we get back the
zero-field Hamiltonian for $q=1$.
From Fig. \ref{fig:phase}, we see that the phase diagrams for $q=2,3$
are similar to the diagram for $q=1\Rightarrow B=0$. However,
there are certain prominent differences -
\begin{enumerate}[label=(\roman *)]
\item Instead of the
gapless phases with 2, 4, 6, and 8 Weyl nodes, we now have gapless
phases with 2q, 4q, 6q, and 8q Weyl nodes. This $q$-fold increase
in the number of Weyl nodes is a consequence of the translation
symmetry of the Hamiltonian in the $y$-direction within a magnetic
unit cell. 
\item For non-zero magnetic fields, there are two 
additional phases which we call W2$'$ and $I'$, which  do not occur
for $B=0$. The spectrum of the phase W2$'$  has  
Weyl nodes in certain bands, while other bands form an LCI. Surface 
states of both types are present. The phase $I'$ is a trivial insulator
in the bulk but has counter-propagating surface states on certain surfaces.
\item Each of the WSM phases gets fragmented and multiple
copies appear with increasing $q$.  We can understand this as
follows. Each of the critical curves (phase boundaries) are determined
by Eq. \ref{eq:gap}, which are inherently $q^{th}$ degree polynomials
in $(M, t_x)$. Thus there are many solutions with increasing $q$.
\item  Despite its increasing complexity with increasing $q$, the phase 
diagram approaches a limit in the $q \to \infty$ limit, with only four 
nontrivial phases W4, W8, LCI, and $I'$ (see Fig. \ref{fig:Largeq}). We will
describe in detail the weak-field limit behavior in Sec. \ref{weakfieldlimit}.
\item The gapped LCI and $I'$ phases take up  regions from the WSM phases
and expand with increasing $q$ and approach a limit when $q$ is large
(see Fig. \ref{fig:Largeq}). Therefore a strong magnetic field can drive
a Weyl  semimetal  to  an insulator with a robust gap ($\sim t_y$, when
$p/q \sim 1/2 \Rightarrow B \sim 10^3$ Tesla, for lattice constant $a \sim 1$ nm)
in our lattice model  (see Sec. \ref{weakfieldlimit}). 
\end{enumerate}

\subsubsection{The LCI phase}
\label{sec:LCIphase}
  As explained earlier, the LCI phase can be 
  thought of as a stack (along x-direction) of 2D Chern layers
  which are tunnel coupled via the hopping $t_x$.  The phase is
  adiabatically connected to the set of disconnected Chern layers in
  the limit $t_x \to 0$. The presence of an external magnetic field
  which is parallel to the Chern layers should not change the LCI
  phase.  In fact, we find that the LCI phase is always present in the
  phase diagram for all $q$. We also note that some parts of the WSM
  phases in the phase diagram of Fig. \ref{fig:phase}a transform to
  the LCI phase in the presence of the magnetic field.

\begin{figure}[ht]
\centering
\includegraphics[width=0.5\columnwidth]{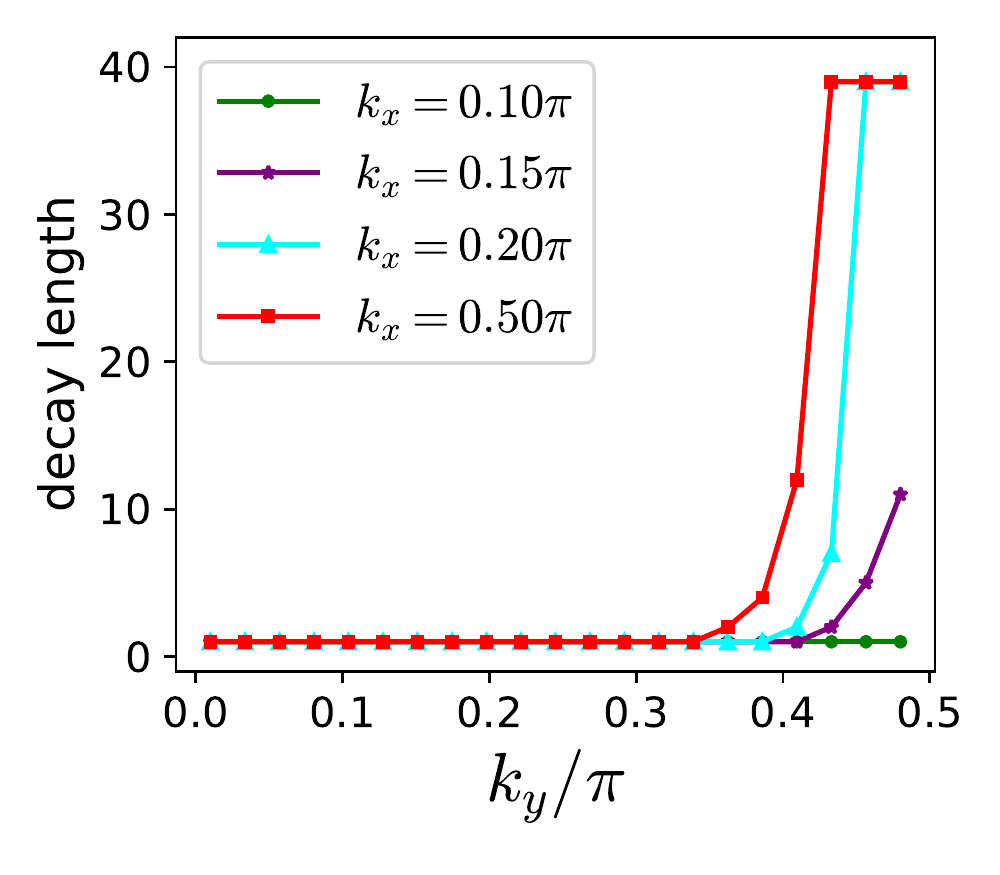}
\caption{Decay length plot of the surface states shown in Figs. \ref{fig:spectrum}e-h, 
for a series of $k_y$ values in between $k_y=0$ and $0.5\pi$. Surface states at 
around $k_y=0$ do not decay into the bulk for any $k_x$ values, but those at 
around $k_y=0.5\pi$ start decaying into the bulk when $k_x$ is taken closer 
to the Weyl point projection  $k_x\approx 0.21\pi$, $k_y=0.5\pi$. When $k_x=0.20\pi$
is very close to Weyl point projection,  decay length of the surface states 
at around $k_y=0.5\pi$ diverges (hits the system size). At $k_x=0.5\pi$, we have
only one type of surface states (but no Fermi arc states associated with Weyl nodes) which
mixes with bulk for $k_y$ values far away from $k_y=0$ (which can be clearly seen in
Fig. \ref{fig:spectrum}h). We compute decay length $l$ by solving  $\ln\left[|\psi(z=0)|^2\right]
=1+\ln\left[ |\psi(z=l)|^2\right]$ numerically, where $\psi(z)$ is the wavefunction of the 
surface states.  }
\label{fig:surface}
\end{figure}

Other than  the LCI  and $I'$ phases, all other phases get fragmented
into multiple copies as $q$ is increased. For example,  there are $2, 3, 5, 
5$, and $3$ copies of W2, W2$'$, W4, W6, and W8 respectively
for $q=3$. The LCI phase, which exists for arbitrary $q$, expands  both along 
$M$ and $t_x$  directions with increasing $q$ (see Fig. \ref{fig:phase}) by
making the system gapped  in a larger region of the  parameter space. It is 
instructive to find the maximum values of $M$ and $t_x$  upto which  the region
occupied by the LCI can reach for a given $q$ and also determine the limits of
these maximum values as $q\to \infty$. So we define the critical values $t^c_x$
and $M^c$ as follows: for $t_x > t^c_x$ the LCI phase  does not exist  for
any $M$, and  similarly for $M > M^c$ the LCI phase does not exist for any $t_x$ 
for a given $q$. For a given $q>1$, it is possible  to obtain the exact expressions 
for these critical values (details in Appendix \ref{appendix:criticalLCI}),

\begin{subequations}
\begin{align}
t^c_x &= t_y ~ 2^{1-1/q}, \\
M^c &= t^{(1)}_z + t^c_x ~\cos{(\pi/2q)} = \left(1 + 2^{1-1/q} \cos{(\pi/2q)}\right) t_y,
\end{align}
\end{subequations}
where we set $t_z^{(1)} = t_y$ in the second equation. For $q=1$, we have  
$t_x^c = t_y$ and  $M^c = 2 t_y$. As $q \to \infty $, these  critical values
approach $ t_x^c = 2 t_y$ and $M^c = 3 t_y$ (see the large $q$ phase diagram in 
Sec. \ref{weakfieldlimit}). So we find that the gapped LCI phase is  
always confined in a finite  region  $M<M^c$, $t_x<t_x^c$ for all values of
magnetic field strength.

\subsubsection{WSM phases and Chern numbers}
Let us now examine the WSM phases in more detail.  We know that in the
zero-field WSM phase, it is possible to understand the existence of
the surface Fermi arc states by computing the Chern numbers
\cite{Savrasov_etal_2011} of the two-dimensional planes that cut the
Fermi arc. A similar analysis here is slightly more subtle. Since in
our model, the Weyl nodes are all along the $k_x$ axis, on the
$k_x$-$k_y$ surface BZ, the projection of all the Weyl nodes are on
the lines $k_y=0$ and/or $k_y= \pi/q$. In general, there are two sets
of Fermi arc surface states on the surface BZ. One set is at $k_y=0$
and the other is at $k_y=\pi/q$ (see Fig. \ref{fig:w2w4arc} for an
illustration). In both cases, the Fermi arcs are dispersionless along
the $k_x$-direction. We can obtain the total Chern number $C(k_x)$ by
integrating the Berry curvature of all the occupied bands of the
two-dimensional magnetic $k_y$-$k_z$ BZ. In our chosen Landau gauge
${\bf A}=(-y, 0, 0)B$, as earlier discussed in the beginning of
Sec. III, we get $C(k_x)=1$ (or $-1$), when $k_x$ belong to the Fermi
arc states at $k_y=0$ (or $k_y=\pi/q$) respectively for all the planes
that cut a single Fermi arc. If the constant $k_x$ plane cuts both the
Fermi arcs at $k_y=0$ and $k_y=\pi/q$, then we get $C(k_x)
=0$. However, a computation of the same quantity in a different Landau
gauge ${\bf A}= B(0, x, 0)$ gives $C(k_x)= q, -q$ and 0 for each of
the three cases above. The reason is that what is physical, and
thus gauge-invariant, is the Chern number per unit length in the
$x$-direction (directly proportional to the Hall current per unit $x$-length).
In the first Landau gauge, the magnetic unit cell is the same size in the 
$x$-direction as the primordial unit cell, while in the second Landau gauge 
the magnetic unit cell is $q$-fold longer in the $x$-direction than the primordial
one. The Chern number per unit length in $x$ is the same in both cases.

\begin{figure}[ht]
\centering
\includegraphics[width=1\columnwidth]{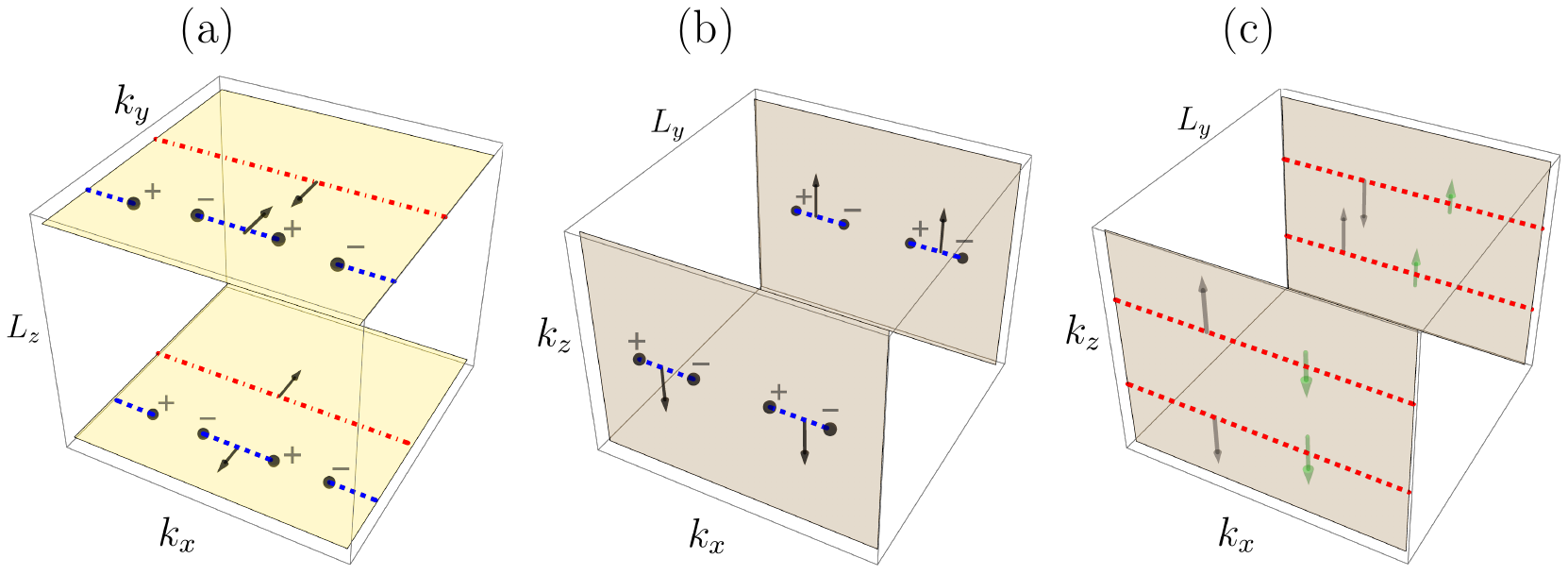}
\caption{The zero-energy Fermi arc 
surface states shown in Fig. \ref{fig:spectrum} in the W2$'$ phase,  are 
depicted here through the cartoon figures (a) and (b), in mixed real and 
momentum space representation. (a) In the $k_x$-$k_y$ surface BZ, there are two
types of surface states: type (i) surface states (shown in blue),
which are the standard Fermi arcs, end at the Weyl point projections ;
type (ii) surface states (shown in red) 
live across the surface BZ. The velocity ${\bf v} = \nabla E(k_x, k_y)$, 
of the particles occupying the type (i) surface states are along the
positive/negative $y$-direction on the top/bottom surface. For
type (ii) surface states, the direction of the velocity gets
reversed. So type (i) and type (ii) surface states are counter
propagating. The Chern number $C(k_x) = 1$, if the constant $k_x$
plane cuts only the type (ii) surface states, else it is zero. 
(b) In the $k_x$-$k_z$ surface BZ, there are 
only type (i) Fermi arc  surface states which end at the Weyl point 
projections. Here we note that the W2$'$ phases which share a boundary
with the $I'$ phase  (see Fig. \ref{fig:phase}c) have two types of surface
states which co-exist on the $k_x$-$k_z$ surface BZ. (c) This cartoon figure
shows the zero-energy  surface states  in the $I'$  phase, in mixed real and
momentum space representation. In the $I'$ phase, there exists localized surface
states only on the  $k_x$-$k_z$ surface BZ.  On both the open crystal
surfaces $y=0$ and $y=L_y$, there are two counter propagating surface states
with green arrows indicating their spin polarization.  Here, the  Chern number
$C(k_x) = 0$, for all $k_x$ planes.}
\label{fig:cartoon} 
\end{figure}

\subsubsection{The phase W2\texorpdfstring{$'$}{'}}

Now let us turn to the W2$'$ phase which did not exist when $
B=0$.  This is a gapless phase with $2q$ Weyl nodes. Its
novelty lies in the fact that it has two types of bulk bands:
those that touch at Weyl nodes, and those that are fully gapped but
carry a nonzero Chern number.
Thus, the W2$'$ is a phase shows the
coexistence of WSM and LCI bands. Consequently, it has two types of
localized surface states - (i) the standard Fermi arc surface states
which are connected to the bulk states at the surface projection of
the Weyl nodes and (ii) surface states 
which are disconnected from the Weyl nodes, and span the magnetic BZ (see
Fig. \ref{fig:spectrum}).  The most important difference between the
two types of surface states is their decay length into the
bulk. The decay lengths of the Fermi arc surface states
diverge at the surface projection of the Weyl nodes whereas the
decay length of the  surface states associated with the LCI bands always
remains finite. In this respect, they resemble the surface 
states of a topological insulator.  The two
types of surface states are also depicted in mixed real and momentum
space representation in Figs. \ref{fig:cartoon}a-\ref{fig:cartoon}b.

The Hall response of the W2$'$ phase is additive between the two types of bands.
The Weyl semimetal Fermi  arc surface states contribute  $k_0(q)q e^2/2\pi h$ to the
Hall conductance per layer~\cite{Burkov_Balents_2011}, while the layer Chern insulator
surface states will contribute $e^2/h$ to the Hall conductance per layer. Since their
surface states  are counter propagating, the total effective Hall conductance per layer is
\begin{equation}
\begin{aligned}\label{eq:Hallcond}
\sigma_{yz}= & \frac{e^2}{h} - \frac{e^2}{2\pi h} q k_0(q)   \\
           = & \frac{e^2}{h}~ \left(1 - \frac{q k_0(q)}{2\pi}\right), 
\end{aligned}
\end{equation}
where $k_0(q)$ is the length of each of the $q$ Fermi arcs connecting the 
$2q$ number of  Weyl point projections on the $k_x$-$k_y$ surface BZ where 
both the LCI  and Weyl semmimetal Fermi arc surface states  exist together
(see also Fig. \ref{fig:cartoon}a). We also note that the length of the 
Fermi arc $k_0(q)$ changes (it decreases) with increasing $q$. So generally 
the Hall conductance in the W2$'$ phase can be tuned  by changing the external
applied magnetic field.

A similiar phenomenon involving  the co-existence of a Weyl semimetal band
and a gapped topological phase has also been recently studied in Ref. \citeonline{rui_etal_2021}.
However, in their case, the WSM  has higher order Weyl nodes, and the topological
bands are part of a higher order topological phase with hinge  states which 
require crystalline symmetry. In our case, we have a simple Weyl semimetal 
and a LCI which co-exist, with no crystalline symmetry needed.


\subsubsection{The phase \texorpdfstring{$I'$}{I'}}
The phase $I'$ is a new topologically trivial phase that appears in the presence
of a magnetic field and
is insulating in the bulk. However, it has zero energy surface states which exist
only on the $k_x$-$k_z$ surface BZ (010 crystal surface). Each of the open
crystal surfaces $y=0$ and $y=L_y$ host a pair of counter-propagating states which
are separated in $k_z$ (see Fig. \ref{fig:cartoon}c). 
A physical way to think about this insulator is as having bulk gapped bands
carrying opposite Chern numbers. The projections of the gapless surface states
corresponding to these bulk bands are naturally counterpropagating, and
separated in $k_z$. If the free surface of the slab is parallel to the
$xy$-plane, $k_z$ is not a good quantum number, and the would-be surface
states hybridize and gap themselves out. However, if the free surface is
parallel to the $xz$-plane then $k_z$ is a good quantum number, and they cannot
gap themselves out.

To support this picture, note that $I'$ always appears adjacent to W2$'$.
From W2$'$, one can obtain $I'$ by the expanding the separation of the Weyl
nodes till the Fermi arc stretches across the BZ before allowing them to annihilate
(see Fig. \ref{fig:cartoon} and caption therein).


\begin{figure}[ht]
\centering
\includegraphics[width=1\columnwidth]{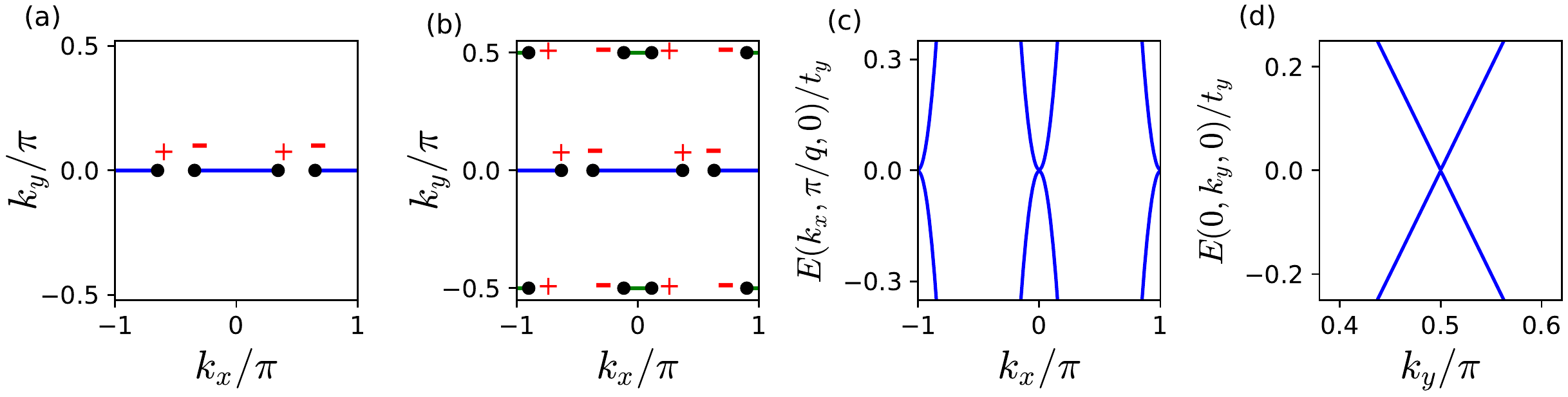}
\caption{Fermi arcs  on the surface BZ $k_x$-$k_y$ in (a) W2 and (b) W4 
for $q=2$.  $M$ is taken to be $2.2t_y$, and  $t_x$ is  tuned  from  $t_x=1.4t_y$
in W2 to $t_x=1.65t_y$ in W4,  which gives rise to a  new pair of Weyl nodes
around $k_x=0$ and $\pm \pi$, at $k_y=\pm \pi/q$, $k_z=0$. The low energy bulk
dispersions   exactly at the phase boundary $M_c=2.2 t_y, t^c_x = 1.56 t_y$, are 
shown in  (c) and (d). The Weyl nodes get  created/annihilated via  a band touching
which is (c) quadratic in $k_x$ and (d) linear in $k_y$. The dispersion along $k_z$ 
is also linear (not shown here). }
\label{fig:w2w4arc} 
\end{figure}

\textcolor{magenta}{\section{Bulk and surface dispersions}}
\label{bulk-surface-dispersions}

In this section, our main aim is to understand how the bulk and the
surface band structures evolve as we vary the parameters to approach
the phase boundaries. We first study  the bulk and surface bands deep 
inside each phase and then analyze how the phase transitions that
occur via the creation/annihilation of Weyl nodes alter the
bulk and surface dispersions. \\

\subsection{Bulk dispersion within the phase}

The spectrum of the $2q\times 2q$ Hamiltonian given in
Eq. \ref{eq:blochH2} has been explicitly derived in Appendix
\ref{appendix:qSol}: 
\begin{align}\label{eq:energydispersion}
E_{n, r}({\bf k}) = r \sqrt{\gamma_n({\bf k}) + (2t^{(2)}_z \sin{k_z})^2 },
\end{align}
with $n=1, 2, 3,..., q$ and $r=(-, +)$. Although we do not have a closed
form expression for $\gamma_n({\bf k})$ (which is non negative for all ${\bf k}$)  
we can numerically compute its
dependence on  the parameters $(M, t_x, t_y, t^{(1)}_z)$ for the flux
$p/q$. We note that in all the gapless  Weyl semimetal phases including
the W2$'$ phase,  the low energy bulk dispersion around each of  the Weyl
nodes ${\bf k}_0$, is linear in all the three directions $k_x, k_y$ and
$k_z$, for all values of $q$. This will change close to the phase boundaries. In
the gapped phases, which are the LCI, $I'$  and  the normal insulator (NI),
the  spectrum has a gap at  the Fermi energy (zero-energy).

\begin{figure}[ht]
\centering
\includegraphics[width=0.8\columnwidth]{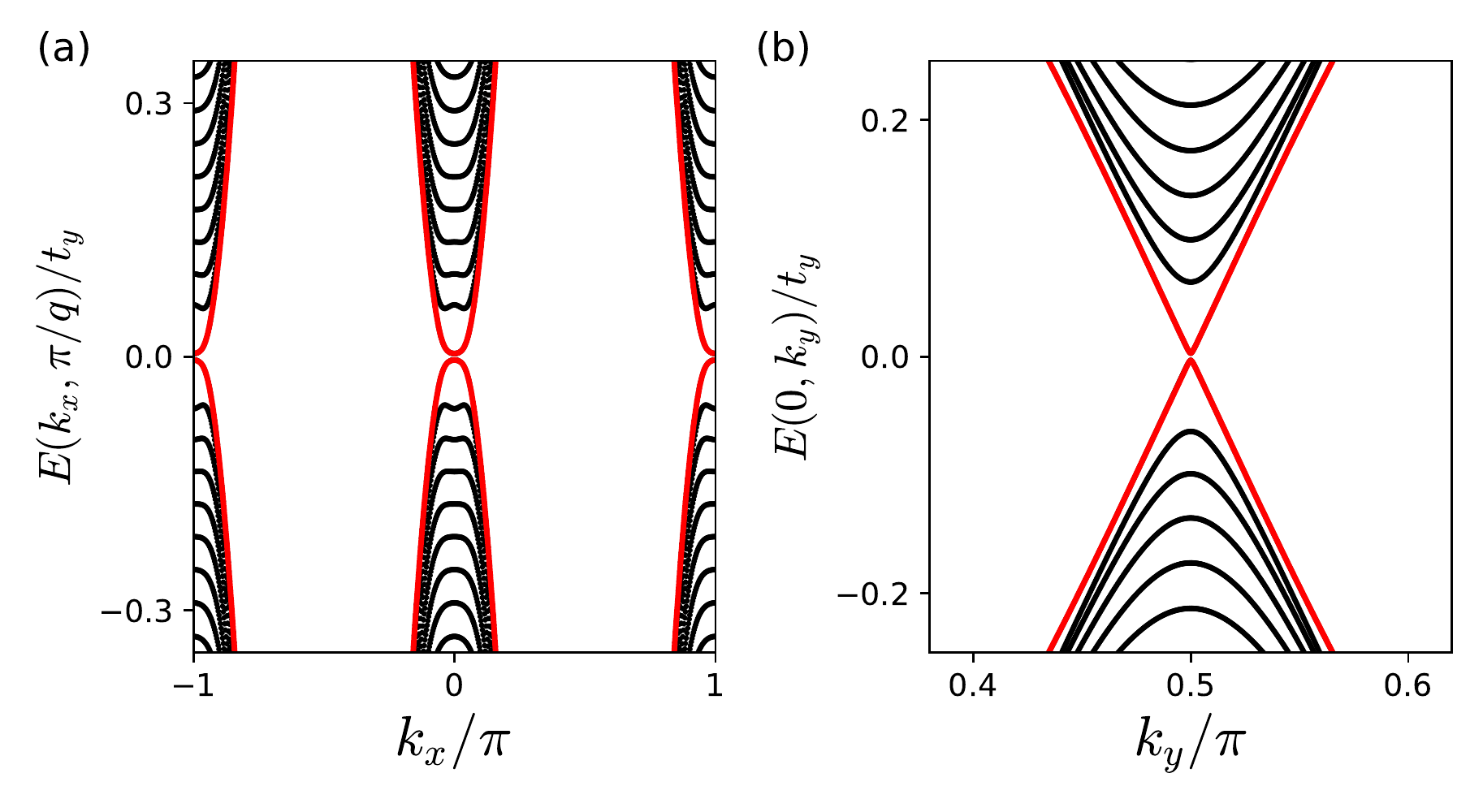}
\caption{The low energy dispersion of the surface states when the
system is a finite slab along the $z$-direction with $L_z=80$, for
$q=2$. The parameters $M$ and $t_x$ are chosen so that the system
is at the boundary between phases W2 and W4. See the caption of
Fig. \ref{fig:w2w4arc} for the choice of parameters. The surface
bands (red) touch (a) quadratically in $k_x$ and (b) linearly
along $k_y$. The low energy surface band touching is similar to  the
low energy bulk band touching (see Fig. \ref{fig:w2w4arc}c-d). The small 
gap at zero energy in the spectrum is due to finite size effects. }
\label{fig:w2w4surface} 
\end{figure}

\subsection{Bulk dispersions at phase boundaries}
\label{sec:BulkDisPB}
The boundaries between the phases in the phase diagram (see for
example, Figs. \ref{fig:phase}b-c), can be between (a) gapped and
gapless phases (b) two gapless phases, and (c) two gapped phases.
Category (c) never occurs in our model, as a gapless phase always
intervenes between two gapped phases, even if its area in
parameter space is exponentially very small. For instance in Fig.
\ref{fig:Largeq}, the intervening
gapless phases are not visible in the phase diagram. The absence
of a direct transition between two gapped phases is a consequence
of the fact that the band structure is continuous in the parameters of
our Hamiltonian. For both categories (a) and (b), the transitions
occur through a band touching which is quadratic in $k_x$, but linear
in $k_y$ and $k_z$. At these transitions, pairs
of Weyl nodes are either created or annihilated. An illustrative case
is depicted in Fig. \ref{fig:w2w4arc}.

\subsection{The dispersion of the surface states within the
phases and at boundaries}

For WSM phases, Fermi arc states are obtained
by diagonalising the Hamiltonian of a finite slab along either the
$z$ (or $y$) directions.  In the surface BZ $k_x$-$k_y$ (or
$k_x$-$k_z$), they are dispersionless along $k_x$ and linear in $k_y$
(or $k_z$). The relevant dispersions in the W2 and W4 phases are shown in
Fig. \ref{fig:w2w4arc}. More details on the dispersion of the surface
states in the W2$'$ phase are shown in Fig.  \ref{fig:spectrum} and
for the $I'$ phase in Fig. \ref{fig:cartoon}c.

As in the bulk spectrum, at phase boundaries, the surface bands
touch quadratically in $k_x$, and linearly in
the remaining direction $k_y$ (or $k_z$) when the system is taken to
be finite along the z (or y) direction. An example is shown in Fig.
\ref{fig:w2w4surface}. One might think that interactions could play
an important role on the 2d surface of the sample because the
combination of the quadratic dispersion in $k_x$ and the linear in
the other direction could lead to a finite density of states (DOS)
at the Fermi energy.  However, it turns out that the low energy DOS
goes as $\sqrt{E}$, and vanishes at zero energy.

\begin{figure}[ht]
\centering
\includegraphics[width=0.6\columnwidth]{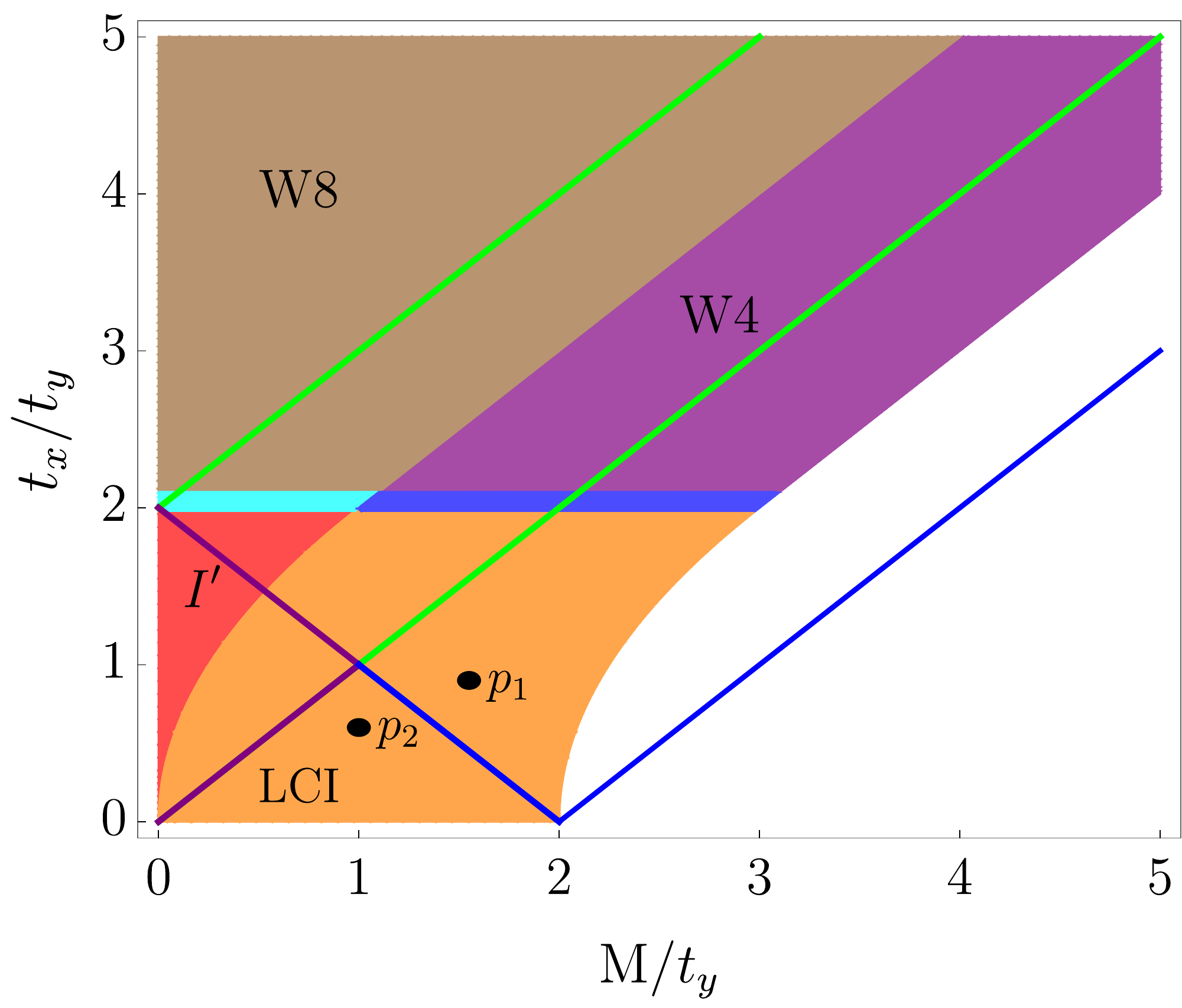}
\caption{The phase diagram for $q=200$. The blue, green and purple straight
lines are the phase boundaries when $ B=0$ (see Fig. \ref{fig:phase}a
for details).  Clearly, for very large $q$, the phase diagram looks much
simpler with just four nontrivial phases  W8 (brown), W4 (purple), 
LCI (orange) and $I'$ (red), and of course the trivial insulator phase. 
We note that the  regions occupied by $I'$, a significant part of  the LCI
and the normal insulator (NI)  above the blue straight line were semimetallic  when  the 
external magnetic field was  zero.
As $q \to \infty$, the widths of the blue region which consists of many
copies of W2 and W2$'$ phases as well as the cyan region which consists
of many copies of W2, W2$'$,  W4, W6, and W8, shrink to zero at $t_x^c =
2t_y$. Along the $M$ direction, the LCI phase expands upto $M^c = 3 t_y$
in the $q\to \infty$ limit.  The critical values $M^c$ and $t_x^c$ are derived in 
Appendix \ref{appendix:criticalLCI} and also discussed in Sec. \ref{sec:LCIphase}.
The points $p_1$ and $p_2$ are discussed in the text.}
\label{fig:Largeq} 
\end{figure}

\textcolor{magenta}{\section{The Weak-Field Limit \label{weakfieldlimit}}}

So far we have studied the phase diagram for small values of $q$. In
this section, we will focus on the phase diagram in the weak magnetic
field limit where $B \to 0$. The $B=0$ or zero flux 
case, corresponding to $q=1$, has been studied earlier in
Sec. \ref{sec:phasediag}. However, the  $q \to \infty$ limit, where the
magnetic flux $\phi \propto 1/q$ goes to zero, is more subtle. Equations
\ref{eq:gapCondition} and \ref{eq:gap}, do not have any simple
behaviour as $q \to \infty$. Nevertheless, the Eqs. \ref{eq:gapCondition}-\ref{eq:gap} 
allow us to construct the phase
diagram for large values of $q$  and consequently
deduce the phase diagram in the limit $q\to \infty$. Surprisingly, the phase
diagram has a simple structure in this limit, with only four nontrivial phases
occupying a significant part of the parameter space - W8, W4, LCI and
$I'$. This  is shown in Fig. \ref{fig:Largeq}. We have checked that the widths
of the blue region which consists of multiple copies of the W2 and W2$'$ phases
and the cyan region which consists of multiple copies of the W2, W2$'$, W4, W6,
and W8, shrink to zero when $q \to \infty$.  Similarly, a thin sliver of the W2$'$
phase lying between the LCI and $I'$ phases shrinks to zero as $q\to\infty$. Clearly,
this phase diagram  looks different from the zero-field phase diagram shown in Fig.
\ref{fig:phase}a (phase boundaries at $B=0$  have been drawn as lines in
Fig. \ref{fig:Largeq} to facilitate the comparison).In particular, regions which
are gapless at $B=0$ appear to be gapped in Fig. \ref{fig:Largeq}. On physical grounds,
we expect the behavior as $B\to0$ to be smoothly connected to that at $B=0$.

To resolve this apparent contradiction, we carry out a more detailed study
of the bulk energy gaps and the low energy dispersions at different points
in the phase diagram. In what follows, we will fix $p=1$, unless stated
otherwise, and examine large values of $q$, choosing  a few 
representative points in the phase diagram in Fig. \ref{fig:Largeq} to show
how the physics as $B \to 0$  is smoothly connected to that at $B=0$. 

\begin{figure}[!ht]
\centering
\includegraphics[width=1\columnwidth]{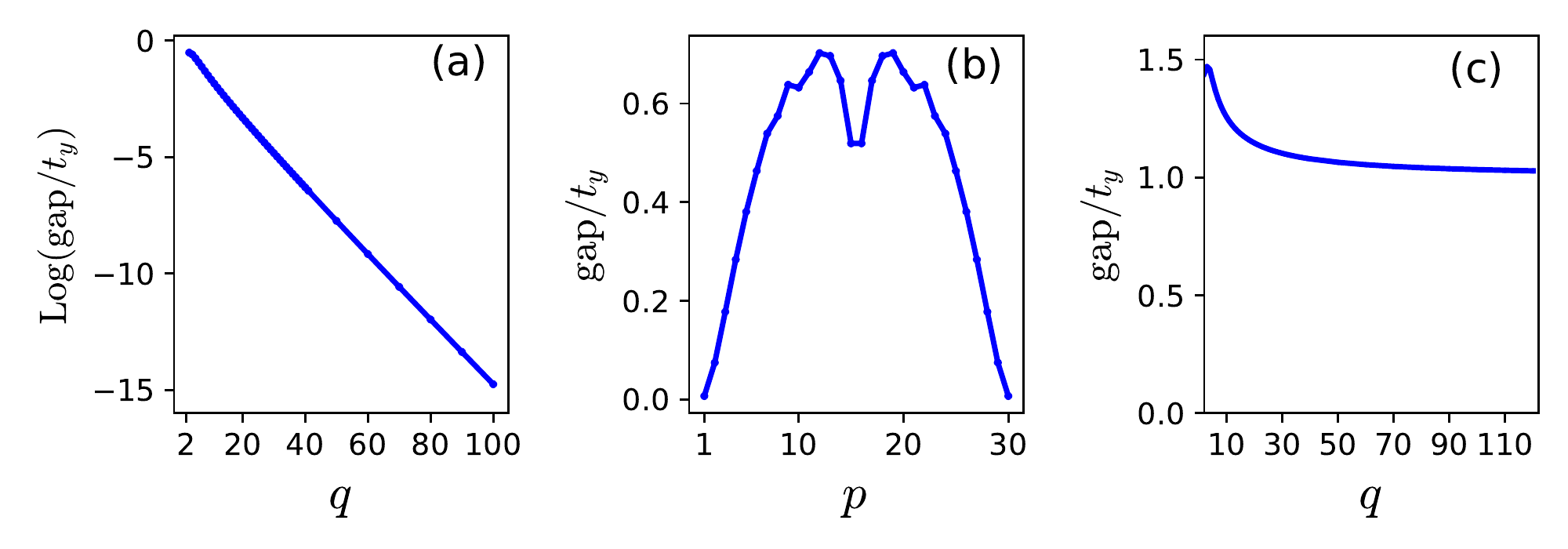}
\caption{(a) The logarithm of the bulk energy gap (computed numerically),
corresponding to a WSM region at $B=0$, plotted as function of $q$, 
for a fixed value of $p=1$, shows the exponential fall  as $e^{-\eta_1 q}$,
with a constant $\eta_1$. The parameters are $M=1.8t_y$, $t_x=0.9 t_y$
correspond  to  the  point $p_1$ in Fig. \ref{fig:Largeq}. (b) The bulk
energy gap is plotted as  a function of $p$ for a fixed  value of $q=31$.
The energy gap first increases with p and then falls again symmetrically
about $p \sim q/2$. (c) Bulk energy gap plotted as function of $q$, 
for a fixed $p=1$, for the point $p_2$  in Fig.  \ref{fig:Largeq},
corresponding to a LCI region at $B=0$. The  parameters are
$M=1.2 t_y$, $t_x=0.4 t_y$. The gap  can be fit to a phenomenological
form $\Delta(q)\simeq \Delta_{\infty} e^{q_0/q}$,
which  does not vanish as $q \to \infty$. }
\label{fig:gapqp} 
\end{figure}

Let us consider an arbitrary point $p_1$ in a region which
is a WSM when $B=0$, but is in the LCI phase when $B \to 0$, 
or at least when $q$ is large. To  check whether the phase is really
semi-metallic or insulating when $q \to \infty$, we need to check
whether the gap vanishes or remains finite in that limit.  To study
this, we compute the bulk energy band gap at $p_1$ as a function of
$q$,  shown in Fig. \ref{fig:gapqp}a. It can be seen that the
gap at zero energy falls exponentially with $q$ at large $q$ for fixed $p=1$.

\begin{figure}[!h]
\centering
\includegraphics[width=0.8\columnwidth]{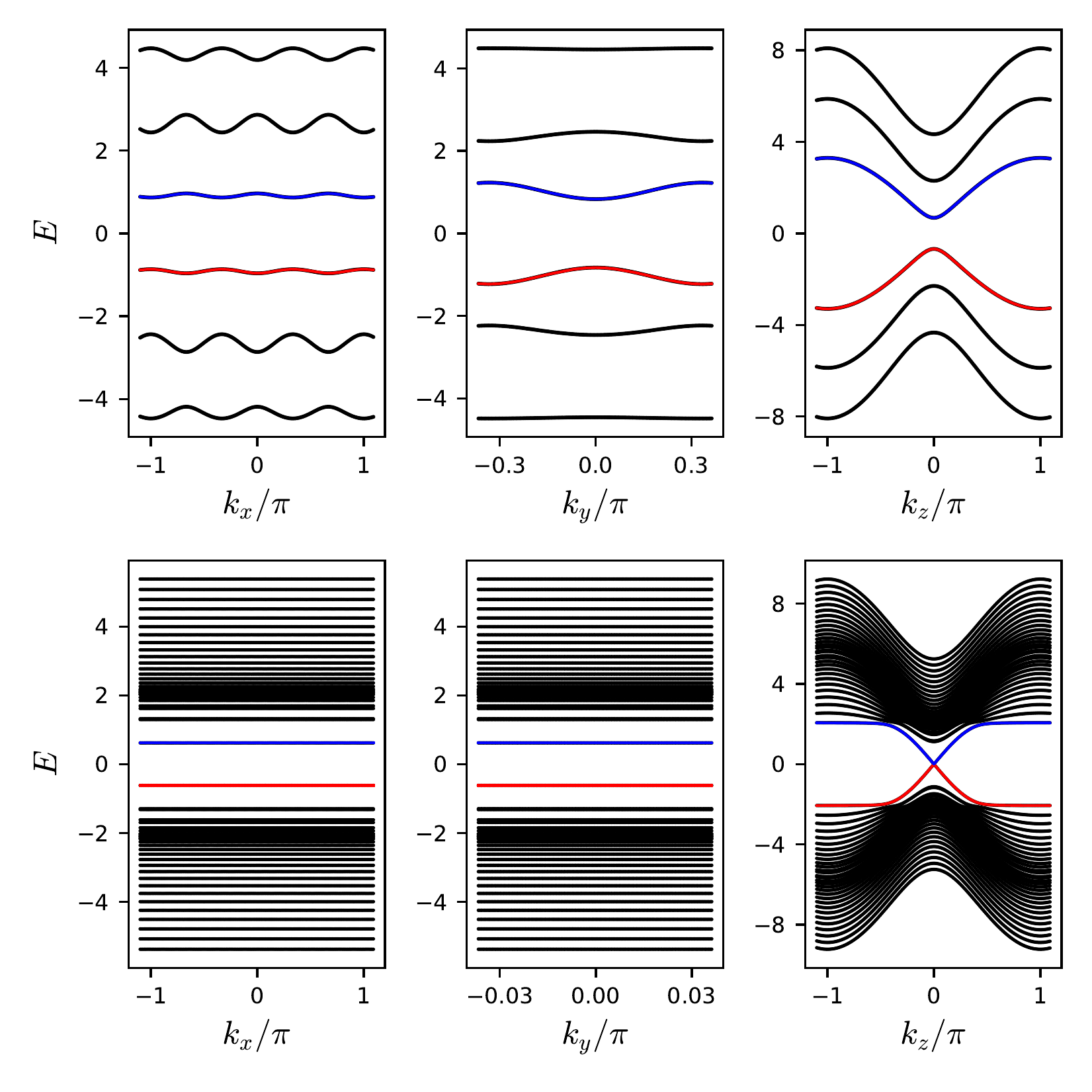}
\caption{The bulk energy spectrum in $k_x$, $k_y$ and $k_z$ directions
for $q=3$ (first row) and $q=30$ (second row). 
The parameters  are $M=1.8$,  $t_x=0.9$, at the  location of the point
$p_1$ in the phase diagram  Fig. \ref{fig:Largeq}. For each of the plots,
one of the momenta is allowed to vary, and the other ones are fixed at
$k_x=0.3\pi$, $k_y=0.2 \pi/q$ and  $k_z=0.1\pi$ appropriately.
The two central  bands are shown in color. Note that the  dispersions in $k_x$ and
$k_y$ become almost flat for $q=30$. Also,  the low energy dispersion as a
function of $k_z$ for $q=30$,  has a very tiny gap, explained further in the
text. }
\label{fig:bandxyz}
\end{figure}

To enable the reader to visualize the change in the spectrum
graphically, the bulk energy spectrum at the  point $p_1$ in
the phase diagram is shown in Fig. \ref{fig:bandxyz} for two 
different values, $q=3$ and $q=30$ (the magnetic field is 
along the $z$ direction as earlier). The dispersions along the 
$k_x$, $k_y$ and $k_z$ directions have been shown explicitly for
both cases. The dispersion in the $k_x$ and $k_y$ directions 
become almost flat for $q=30$, corresponding to the semiclassical 
Landau levels. Furthermore, we note that already for $q=30$,
the two central bands come close together at $k_z=0$ with a very
tiny gap, approaching the linear dispersion, in the field direction 
of the $n=0$ Landau levels in the semiclassical limit. In fact, even  
for $q=30$, the low energy dispersion for small $k_z$ in our lattice model 
looks  almost identical to the dispersion of Weyl fermions in the continuum 
model\cite{Goerbig_etal_2008,Burkov_Hook_2011,Serguei_Marcello_Goerbig_2016,Li_Roy_DasSharma_2016}. In the $q \to \infty$ limit, clearly, 
this is the W2 phase (there are only  two linearly dispersing $n=0$ Landau
levels). This is true even though, for any finite $q$, the phase is gapped
and is in the LCI phase. It is just that the gap is exponentially small in
$q$. We have also checked that when the point $p_1$ is moved to regions 
where the zero field case is in the W4 or W6 phase, the bands come close
together both at $k_z=0$ and $k_z=\pi$, signifying that in the $q\to
\infty $ limit, they evolve to Weyl nodes at both $k_z=0,\pi$. \\

Now let us consider another point $p_2$ in a region which is a
LCI both when $B=0$ and  $B\to 0$.  Here, when we compute the bulk
energy gap as a function of $q$, we find that it behaves roughly as
$\Delta\simeq \Delta_{\infty}e^{q_0/q}$, which does not vanish as
$q\to\infty$. This is shown in Fig. \ref{fig:gapqp}c.  Further, we
also note that in regions which are gapless semi-metals, both when
$B=0$ and $B\to 0$, the low-energy dispersion of the $B\to 0$ phase
approaches that of the $B=0$ phase.  We hence conclude that
although naively the $q\to \infty$ phase diagram looks different
from the zero-field ($q=1$) phase diagram, the physics they describe,
in terms of their bulk energy gaps and their low-energy
dispersions, is smoothly connected via the semiclassical description.

We end by noting the
behaviour of  the bulk energy gap as a function of
$p$ and $q$, which is shown in Fig. \ref{fig:gapqp}. 
The energy gap decreases with $q$ for a fixed $p$  and  increases with 
$p$ for a fixed $q$, and  the maximum gap $\sim t_y$  occurs when $p/q \sim 1/2$.
Thus for large $q$ values and when $p/q \sim 1/2$, a very strong 
magnetic field $B = \frac{p}{q}\frac{\phi_0}{a^2} \sim 10^3$ Tesla (if lattice 
constant $a=1$ nm ) can drive a Weyl-semimetal to an insulator
by annhilating the Weyl nodes with gap $\sim t_y$ in a robust region of the parameter
space.At a large magnetic field, annihilation of Weyl nodes is expected to open a gap
when the   inverse magnetic length $\sqrt{eB/\hbar} \sim k_0$,  the separation between the Weyl 
nodes. This has been observed also in experiments by measuring the resistivity of  Weyl materials
TaP \cite{CL_Zhang_etal_2017} and TaAs \cite{Ramshaw_etal_2018}  in high applied  magnetic
field.


\textcolor{magenta}{\section{Summary and Outlook}}

In summary, we have studied a time-reversal broken WSM in the presence
of a commensurate orbital magnetic field with minimal crystalline symmetry. 
We have considered the
case where the direction of the magnetic field is normal to the
line joining the Weyl nodes in the absence
of a magnetic field, which we denote as the $x$-direction. In the presence 
of a $p/q$ flux per unit
cell, we have obtained the phase diagram in the parameter
space of the onsite mass $M$ and the $x$-direction hopping $t_x$.
Setting two of the five independent hoppings equal ($|t_y^{(1)}|=|t_y^{(2)}|$)
allows us to solve for the entire phase diagram analytically  for arbitrary $p,q$. 

We find that  the phase diagram contains WSM phases hosting $2q$,
$4q$, $6q$, and $8q$ Weyl nodes. These phases occur in multiple 
copies for nonzero flux, with the number of copies depending on $q$. 
The gapped LCI phase  also exists for arbitrary $q$. There are two
additional phases, which we call W2$'$ and $I'$, that appear in the
phase diagram only at nonzero flux. The phase W2$'$ has a  gapless bulk
spectrum  with $2q$ Weyl nodes, but has additional gapped bulk bands which
carry nontrivial Chern number in the $k_y$-$k_z$ plane at fixed $k_x$. This
phase displays a coexistence of Weyl semimetal and layered Chern insulating
behavior. In accordance with the bulk-boundary correspondence, the W2$'$
phase has Fermi arc states at the surface, as well as surface states required
by the layered Chern insulator. These two types of surface states in the W2$'$
phase are counter-propagating. The phase $I'$ is fully gapped in the bulk but
hosts a pair of counter-propagating surface states on the $xz$ crystal surface,
but none on the $yz$ surface.

The fact that we can analytically obtain the entire
phase diagram for arbitrary $p,q$ enables us to systematically study
the weak-field limit $q\to\infty$, and to smoothly relate it to
the $B=0$ limit. Formally, for $p=1$ and any
large but finite $q$, the phase diagram looks quite different from the
zero-field case $p=q=1$. However, an examination of the spectral gap
at zero energy reveals the way the limit should be taken
physically. In regions of parameter space where the zero-field case is
fully gapped, the $q\to\infty$ gap remains finite when the limit is
taken, whereas in the regions of parameter space where the $q=1$ case
is gapless and the large $q$ case appears gapped, the gap of the
latter vanishes exponentially as a function of $q$.

Let us briefly consider the stability of these phases to
time-reversal symmetric potentials, either commensurate with the
lattice, or arising from quenched disorder. Focusing on commensurate
potentials (of very small amplitude compared to the bandwidth),
provided the separation between the bulk Weyl nodes is commensurate,
any WSM phase can be gapped out in the bulk by an appropriate
periodic potential. This will naturally gap out the corresponding
surface Fermi arcs as well. Coming to the W2$'$ phase, such a
periodic potential would gap out the Weyl nodes, but cannot destroy
the Chern numbers of the fully gapped bands. Thus, we conclude that
the W2$'$ phase is unstable to becoming a simple LCI. Similarly, the 
$I'$ phase becomes a trivial insulator when subjected to a small
periodic potential of the appropriate period, which induces matrix
elements between the counter-propagating surface modes and gaps them
out.

Extending our discussion to quenched disorder, two lines of
argument have been explored in the literature. It is known 
that the WSM is perturbatively stable to disorder in the
renormalization group sense \cite{Fradkin_1986_I,
Fradkin_1986, Goswami_etal_2011, Altland_2015,Roy_etal_2018,
Kobayashi_etal_2014,Sbierski_etal_2014,Louvet_Carpentier_Fedoreko_2016}.
In this picture, the WSM undergoes a transition to a diffusive metal at
a nonzero critical disorder strength. Along a different line of
reasoning, however, the nonperturbative effects of large but rare
Griffiths regions have been argued to destroy the WSM at arbitrarily
weak disorder \cite{Rahul_2014, Pixley1_2016, Pixley2_2016, Pixley3_2017},
and make it a diffusive metal at the longest length-scales. If the first
scenario prevails, the WSM phases uncovered in this work (and, significantly,
the W2$'$ phase) will be stable to weak disorder. However, if the second
scenario is proven to hold generically, the WSM phases and the W2$'$ phases
will be destroyed immediately for arbitrarily weak disorder. This is because
a diffusive metal will destroy the quantization of the Hall conductance
immediately. However, the LCI and the $I'$ phases, being fully gapped in
the bulk, are expected to be stable to arbitrary weak disorder. \\

Let us comment briefly on potential experimental realizations
of the physics uncovered here: There is a theoretical proposal for the simplest
Weyl semimetal with only two Weyl nodes, based on inserting magnetic layers of
Mn into a layered topological insulator such as $\mathrm{Bi}_2\mathrm{Sb}_3$
\cite{Kong_etal_2010}. According to theory, the material should be a WSM with 
two Weyl nodes when the Mn layers are ferromagnetic \cite{Lei_Chen_MacDonald_2020}.
Clearly, very large magnetic fields are  needed to realize even $q\simeq100$ in
this realization. In a different direction, there are several proposals for
realizing a Weyl semimetal in an optical lattice
\cite{Zhou_etal_2016,Tena_etal_2015,Xu_Zhang_2016}. Very recently one such proposal
has been also experimentally realized \cite{Wang_etal_2021}. Since
orbital fields can be imposed in a frequency driven optical
lattice \cite{Eckardt2017,Monika_etal2018}, approaching commensurate fields seems more
achievable in this realization.

Many open questions remain. An important one is the effect of interactions on the
phase diagram. At $B=0$, interactions have been conjectured to drive the WSM
into numerous phases, including charge density wave \cite{Wang-Zhang2013,Youetal2016,
Weideretal2020,Thakurathi2020}, superconducting \cite{Weietal2014}, Mott insulating
\cite{Morimoto2016}, and even fractionalized \cite{Hotzetal2019,Sagietal2018} phases.
Renormalization group analyses have also been carried out keeping all symmetry-allowed
interactions \cite{Maciejko_Nandkishore_2014,Roy_Goswami_etal_2017}. As we have shown,
near a quantum phase transition between different WSM phases, the spectrum near zero
energy becomes quadratic in one direction, remaining linear in others. This would
suggest a greater instability to certain types of interactions near the phase boundaries.

Our work could also be extended in a different direction. Our starting
point in this study was a time-reversal broken WSM;  adding  the orbital
magnetic field did not change this. However, we did find  extra phases W2$'$ and
$I'$ which appear only in the presence of a magnetic field. It would
be of interest also to study time-reversal symmetric, inversion
broken, Weyl semi-metal in a magnetic field, where the introduction of
the magnetic field would introduce time-reversal breaking as well.
We look forward to answering these and other questions in future work.


\textcolor{magenta}{\section{Acknowledgement}}
AD was supported by the German-Israeli Foundation (GIF) Grant No.
I-1505-303.10/ 2019 and the Minerva Foundation. AD also thanks Dean of Faculty
fellowship Weizmann Institute of Science, Israel, Israel planning and
budgeting committee,  and Koshland Foundation for financial support.
SR and GM would like to thank the VAJRA scheme of SERB, India for its
support. GM is grateful for partial support from the US-Israel Binational
Science Foundation (Grant No. 2016130) and the Aspen Center for Physics 
(NSF grant PHY-1607611) where this was completed.\\


\appendix

\section{Symmetries of the lattice model}
\label{appendix:symmetryanalysis}

We have considered a minimal two-band lattice model of Weyl semimetals,  Eq. \ref{eq:realspaceH},
which breaks time reversal symmetry but  keeps the inversion symmetry. Since the 
model involves pseudospin, the time reversal operation is just complex conjugation. 
For the case of isotropic hoppings, the symmetries  of this model can be found  in 
Ref. \citeonline{Abdulla_etal_2021}. 
Now let us look at in more detail at  the crystalline symmetries which our anisotropic  model has: 
\subsection{Rotation:}
It has a single two-fold rotation symmetry only about the $x$-axis and the symmetry transformation 
($C_{2x}$) is given by
\begin{equation}
\begin{aligned}
    C_{2x}~ c_s(n_x,n_y,n_z)C_{2x}^{-1} = (\sigma_x)_{ss'} ~ c_{s'}(n_x, -n_y,-n_z).
\end{aligned}
\end{equation}
It can be easily checked that this symmetry operation leaves the Hamiltonian in
Eq. \ref{eq:realspaceH} invariant i.e. $C_{2x} H C_{2x}^{-1} = H$. 

\subsection{Mirror reflection:} 
The model is symmetric under the following mirror reflections about the
$yz$ ($M_x$), $xz$ ($M_y$) and $xy$ ($M_z$) planes:

\begin{equation}
\begin{aligned}
    M_x~ c_s(n_x,n_y,n_z)M_x^{-1} = (\sigma_0)_{ss'} ~ c_{s'}(-n_x, n_y, n_z), ~~ M_x i M_x^{-1} = i
\end{aligned}
\end{equation}

\begin{equation}
\begin{aligned}\label{eq:Mysymmetry}
    M_y~ c_s(n_x,n_y,n_z)M_y^{-1} = (\sigma_x)_{ss'} ~ c_{s'}(n_x, -n_y, n_z), ~~ M_y i M_y^{-1} = -i
\end{aligned}
\end{equation}

\begin{equation}
\begin{aligned}
    M_z~ c_s(n_x,n_y,n_z)M_z^{-1} = (\sigma_0)_{ss'} ~ c_{s'}(n_x, n_y, -n_z), ~~ M_z i M_z^{-1} = -i. 
\end{aligned}
\end{equation}
All the  above three symmetry transformations $M_x$, $M_y$ and $M_z$ leave the Hamiltonian in
Eq. \ref{eq:realspaceH} invariant i.e. $M_{\mu} H M_{\mu}^{-1} = H$, $\mu=x, y, z$. We note
that the symmetry operations given by $M_y$ and $M_z$ are not pure mirror reflections, rather 
they  can be thought of as a product of mirror and time reversal operations. On the other hand,
the two fold rotation $C_{2x}$ about $x$ axis and the mirror reflection $M_x$ about the $yz$ plane
together give the  inversion operation, and therefore the Hamiltonian is also inversion symmetric.

\subsection{Particle-hole transformation:}
We find that the following particle-hole transformation, which also  includes the  mirror reflection  about the
$xz$ plan, 
\begin{equation}
\begin{aligned}\label{eq:phsymmetry}
    P_x c_s(n_x,n_y,n_z)P_x^{-1} = (\sigma_x)_{ss'} ~ c^{\dagger}_{s'}(n_x, -n_y, n_z), ~~  P_x i P_x^{-1} = i
\end{aligned}
\end{equation}
takes the Hamiltonian $H$ in Eq. \ref{eq:realspaceH} to $-H$ i.e. $P_x H P_x^{-1} = -H$. Therefore 
this symmetry forces the spectrum of $H$ to be symmetric about zero energy. \\

Now we check on which of  these symmetries  survive in the presence of an applied uniform external 
magnetic field along the $z$-direction. In presence of the  magnetic field, the $x$-hopping $T_x=t_x\sigma_x$
picks
up a phase factor $\exp(-i2\pi y \phi/\phi_0)$, which breaks all the symmetries,  except  the mirror
reflection $M_y$ about the $xz$ plane (Eq. \ref{eq:Mysymmetry}) and the  particle-hole transformation 
$P_x$ (Eq. \ref{eq:phsymmetry}). Therefore the symmetry realized  by $P_x$ survives even in the presence of the  magnetic field,  and ensures that  the spectrum   is symmetric about  zero energy. 


\section{Fermi arc surface states at zero and finite commensurate fields }
\label{appendix:sim}

The pseudospin polarization of the Fermi arc surface states changes its
polarization character as we go from  zero field to finite commensurate
fields for both the WSM and the LCI phases. At zero field, the  surface
states are polarized in pseudospin along the $z$-direction. The right 
moving surface states of a  slab finite along $z$-direction are localized
at the top surface $z=L_z$  and they are polarized along the positive
$z$-direction with pseudospin $s = \uparrow$. In contrast, the left
moving surface states are localized at the bottom surface $z=0$ and they
are  polarized along the negative $z$-direction with  pseudospin $s= \downarrow$.

\begin{figure}[H]
\centering
\includegraphics[width=1\columnwidth]{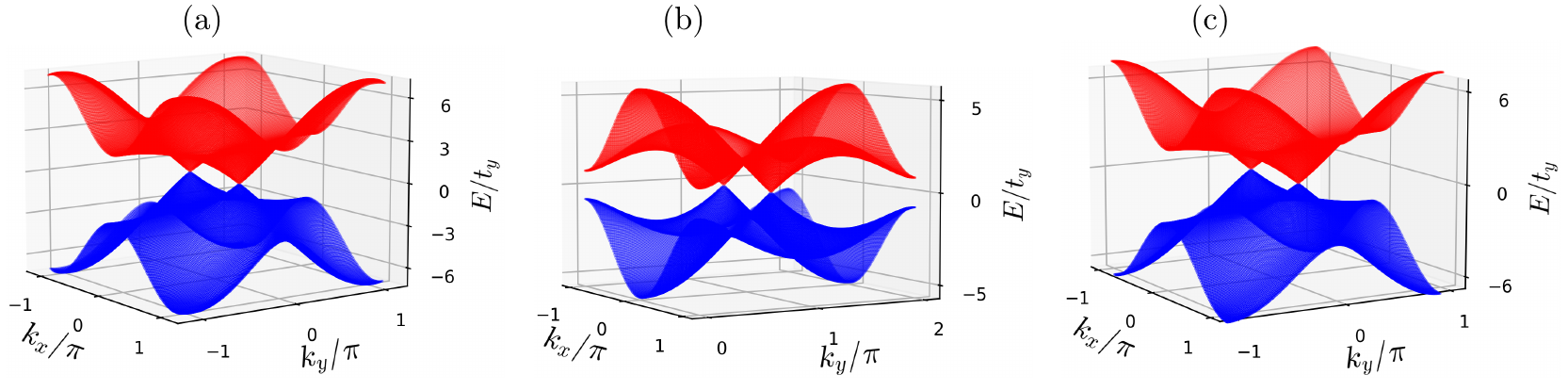}
\caption{Energy spectrum of the zero field model showing Weyl nodes 
with linear dispersion along all directions near zero energy. (a) There
are two Weyl nodes in the W2 phase which appear in  the $k_z=0$ plane.
There exist  four Weyl nodes in the W4 phase: two of them occur in the  (b)
$k_z=0$ plane and the other two in the (c) $k_z=\pi$ plane}
\label{fig:q1band3d} 
\end{figure}

\begin{figure}[H]
\centering
\includegraphics[width=0.8\columnwidth]{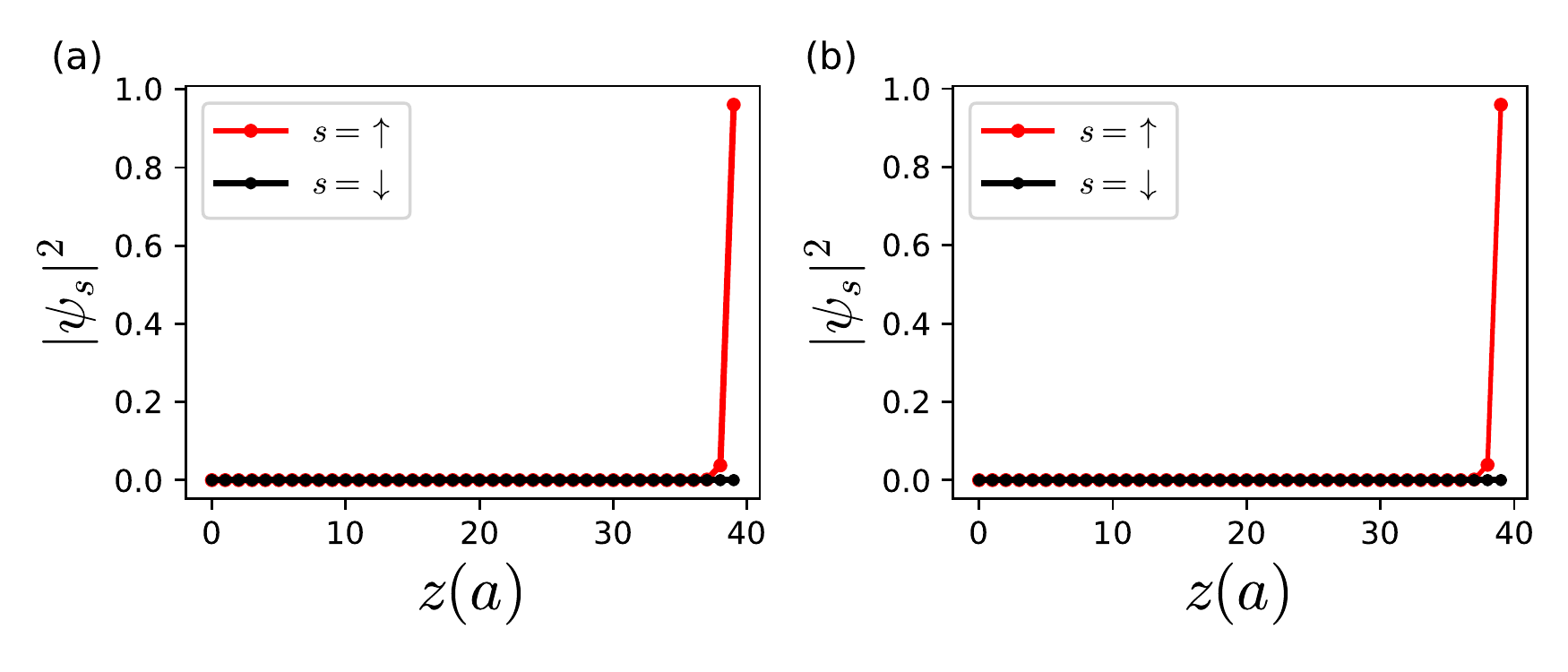}
\caption{Eigenstate plots of the right moving Fermi arc surface states for the zero
field  (${\bf B }=0$) case for a finite slab along the $z$-direction with lattice sites  $L_z=40$.
(a) This is in the LCI phase: the right moving surface  states are localized on the
top surface $z=L_z$ and are polarized along $z$-direction with  pseudospin $s= \uparrow$.
(b) This is in the W2 phase: the right moving Fermi arc states are
localized on the top surface $z=L_z$ and they are polarized along $z$-direction 
with pseudospin $s=\uparrow$.}
\label{fig:q1surface} 
\end{figure}

\begin{figure}[H]
\centering
\includegraphics[width=0.8\columnwidth]{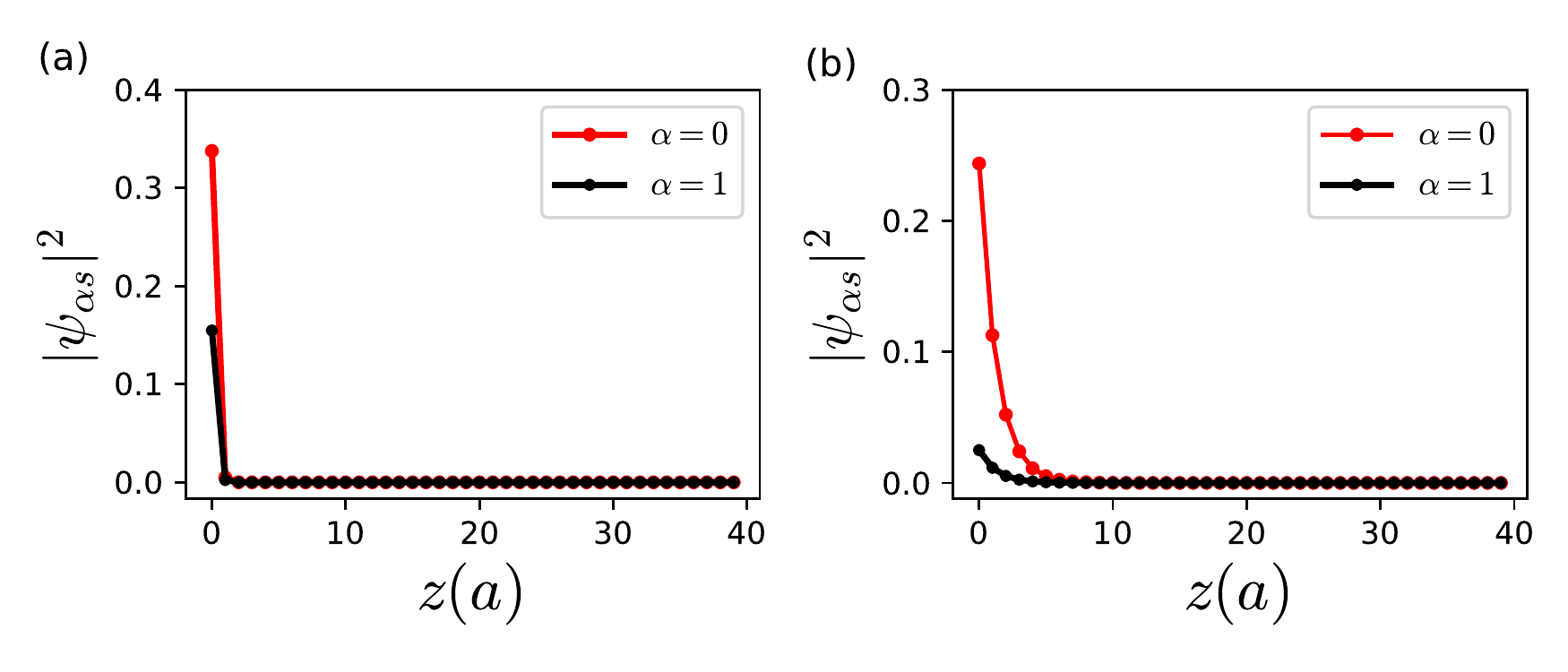}
\caption{Eigenstate plots of the  right moving Fermi arc surface states in
the presence of a finite commensurate field ($B \neq 0$) with $q=2$,
 for a finite slab along the
$z$-direction with lattice sites  $L_z=40$.  Here $\alpha=0,1$ refer to the 
coordinate of the  lattice sites inside the  magnetic  unit cell. (a) This is in the 
LCI phase: the right moving Fermi arc states are localized on the bottom surface 
$z=0$ and  both the pseudospin components
have the  same amplitude. (b) This is in the W2 phase: the right moving Fermi arc
states are localized at the bottom surface $z=0$ and both the pseudospin components 
have the same amplitude.}
\label{fig:q2surface} 
\end{figure}

A few eigenstates plots are shown in 
Fig. \ref{fig:q1surface} for this zero field case. When a magnetic field
is switched on in the  $z$-direction, surface states in the $k_x$-$k_y$
surface BZ of both  the WSMs  and  LCI phases   alter their polarization
from the  $z$-direction to the $x$-$y$ plane. Now the top surface hosts
left moving surface states and the bottom surface hosts right moving
surface states with amplitude being the same for both the $s =\uparrow$
and $s = \downarrow$ pseudospins. A few eigenstate plots are shown
in Fig. \ref{fig:q2surface} for non zero magnetic field with $q=2$.

\section{Zeros of Bloch-Hofstadter Hamiltonian for arbitrary \texorpdfstring{$q$}{q}}
\label{appendix:qSol}
In this section, we will go through a detailed discussion  of how we obtained  the
topological phase boundaries, the critical values and the phase diagram in the
presence of an  arbitrary commensurate magnetic field. We begin by
writing the Hofstader Hamiltonian $h_{\phi}({\bf k})$ in Eq.
\ref{eq:blochH2} as  $h_{\phi}({\bf k}) = \Psi^{\dagger} ~
\Tilde{h}_{\phi}({\bf k}) ~ \Psi$, with  $\Psi =(\psi_{\uparrow},
\psi_{\downarrow})^T$, where $\psi_{s} = \left(c_{0,s}({\bf k}),
c_{1,s}({\bf k}), ..., c_{q-1,s}({\bf k})\right)^T$
and $s \equiv \left(\uparrow, \downarrow\right)$. The matrix
$\Tilde{h}_{\phi}({\bf k})$ is given by
\begin{align}
\Tilde{h}_{\phi}({\bf k}) = \begin{pmatrix}
{\bf A} & B \\
{\bf C} & {\bf D}
\end{pmatrix}
\end{align}
where all the blocks are of same dimension $q\times q$ and are given by 
\begin{subequations}
\begin{align}
{\bf A} =&~ -{\bf D} = 2t^{(2)}_z \sin{k_z} {\bf I}_q ~~{\rm and} \\
B=&~ {\bf C}^{\dagger} = \begin{bmatrix}
m_0 & -u & 0 & 0 & ... & -v e^{ik_y q} \\
-v & m_1 & -u & 0 & ... & 0 \\
0 & -v & m_2 & -u & ... & ... \\
.. & .. & .. & .. & .. & .. \\
0 & 0 & ... & -v & m_{q-2}& -u \\
-ue^{-ik_y q}& 0 & ... & ...& -v & m_{q-1}~. \\
\end{bmatrix}
\end{align}
\end{subequations}
Here ${\bf I}_q$ is the identity matrix of dimension $q \times q$, $ u
= t^{(1)}_y - t^{(2)}_y$, $v =t^{(1)}_y + t^{(2)}_y$,  $m_{\alpha}
= f^{\alpha}_1({\bf k}) = 2\left( M - t^{(1)}_z \cos{k_z} - t_x
\cos{\left(k_x + \frac{2 \pi  p}{q}\alpha\right)}\right)$, and $\alpha\in[0,q-1]$ is the
sublattice index. The eigenvalues $\lambda$ satisfy
\begin{align}\label{eq:det1}
\textrm{det} \begin{bmatrix}
{\bf A} - \lambda {\bf I}_q & B \\
{\bf C} & {\bf D} - \lambda {\bf I}_q \\
\end{bmatrix} 
= 0. 
\end{align}
Clearly the matrix  ${\bf C}$ commutes with both
$\tilde{{\bf D}} = \left({\bf D} - \lambda {\bf I}_q\right) = -
\left(\lambda + 2t^{(2)}_z \sin{k_z}\right) {\bf I}_q$ and $\tilde{{\bf
    A}} = \left({\bf A} - \lambda {\bf I}_q\right) = - \left(\lambda -
2t^{(2)}_z \sin{k_z}\right) {\bf I}_q$. The left hand side of
Eq. \ref{eq:det1} is identically equal to $\textrm{det}\left({\bf
  \tilde{A} \tilde{D}} - B {\bf \tilde{D}}^{-1} {\bf C} {\bf
  {\tilde D}}\right)$. Using the commutation of
${\bf C}$ and ${\bf {\tilde D}}$, Eq. \ref{eq:det1} can be simplified
to
\begin{align}\label{eq:det2}
\textrm{det}\left(\gamma {\bf I}_q - {\bf B B}^{\dagger}\right) = 0, 
\end{align}
where we have used ${\bf \tilde{A} \tilde{D}} = \gamma {\bf I}_q $, $\gamma =
\lambda^2 - \left(2t^{(2)}_z \sin{k_z}\right)^2 $ and ${\bf C} = B^{\dagger}$. So the
eigenvalues are given by 
\begin{align}\label{eq:eigenvalue}
\lambda = \pm \sqrt{ \gamma + \left(2t^{(2)}_z \sin{k_z}\right)^2 }.
\end{align}
The spectrum (we have suppressed the momentum argument ${\bf k}$ of $\gamma$
and $\lambda$ for notational simplicity) is obviously  particle-hole symmetric,
which is a consequence of the particle-hole symmetry $P_x$ of 
the Hamiltonian discussed in Appendix \ref{appendix:symmetryanalysis}. 
Eq. \ref{eq:det2} implies that $\gamma$ is the eigenvalue
of the positive semi-definite matrix ${\bf B B}^{\dagger}$ and so it must
satisfy $\gamma\geq 0$. Therefore, for  the spectrum to be gapless (which means
$\lambda=0$), we must have $\gamma=0$ and $\sin{k_z} = 0$. The only
possibilities are $k_z=0$ and $\pi$ at which energy gap can close in
the $k_z$ direction. 

Since $\gamma=0$ for energy spectrum to be gapless, we get the following
determinant vanishing condition from Eq. \ref{eq:det2}, 
\begin{align}\label{eq:det3}
\textrm{det}(B) = 0,
\end{align}
where we have used the identity $\textrm{det}({\bf B B}^{\dagger}) =
\textrm{det}(B) \textrm{det}(B^{\dagger})$. To get the explicit
gap closing condition in terms of the parameters of the theory, we must
compute the determinant of the $q \times q$ matrix $B$. Looking at
the matrix $B$, we find that it is particularly simple to calculate
determinant of $B$ if either $ u = t^{(1)}_y - t^{(2)}_y$ or $v
=t^{(1)}_y + t^{(2)}_y$ is zero ($B$ then becomes almost a lower/upper
triangular matrix). This can happen only when $|t^{(1)}_y| = |t^{(2)}_y|$. We
choose $t^{(1)}_y = t^{(2)}_y$ ($=t_y$ say) such that $u=0$ as
discussed in the main text, and  we use $t_y$ as a  unit of energy. The other
choice $t^{(1)}_y = -t^{(2)}_y$  which makes $v=0$ will be equally valid and will not make 
any difference to our final results. Now we can expand the determinant of $B$
(setting $u=0$) about the first row to get
\begin{align}\label{eq:det4}
\textrm{det}(B) = & \left\{ \prod_{\alpha=0}^{q-1} m_\alpha \right\} + (-1)^{q+1} (-v
e^{ik_y q}) (-v)^{q-1} \nonumber \\
= & \left\{\prod_{\alpha=0}^{q-1} m_\alpha \right\} - e^{ik_y q} v^q.
\end{align}
For $\textrm{det}(B)=0$, the imaginary part $-v^q \sin{qk_y}$
and the real part must vanish. This implies  that the energy gap can 
close only at $k_y=0$ and
$\pi/q$, along the $k_y$ direction (recall $0 \leq k_y \leq 2\pi/q$). For
the real part  of $\textrm{det}(B)=0$ to vanish we must have 
\begin{align}\label{eq:det5}
\prod_{\alpha=0}^{q-1} m_\alpha = (-1)^{\mu} v^q = (-1)^{\mu} 2^q t_y^q ,
\end{align}
where $\mu$ takes values 0 and 1 corresponding to the closing of the gap at
$k_y=0$ and $\pi/q$ respectively. We have used $v = t^{(1)}_y + t^{(2)}_y =
t_y + t_y = 2t_y$ in the above expression. Recall  that all the $k_x$
dependence is in the $m_\alpha = 2\left(M - (-1)^{\nu} t^{(1)}_z - t_x
\cos {\left(k_x + \frac{2\pi p}{q} \alpha\right)} \right)$.  Here $\nu$
takes values 0 and 1 which correspond to closing of the  gap at $k_z=0$ and
$\pi$ respectively. Equation \ref{eq:det5} gives us the final set of conditions
which are to be satisfied by all the parameters of the theory for the
energy spectrum to be gapless and thus  completely
determines the phase boundaries of the topological phase diagram. 

It is possible to simplify the left hand side even further. We will use the
following identity for the finite product \cite{zwillinger2012crc}
\begin{align}\label{eq:productFormula}
\prod_{\alpha=0}^{q-1} \left(2g + 2\cos{\left(k_x + \frac{2\pi p}{q} \alpha\right)} \right) = 
2 \left[ T_q(g) + (-1)^{p+q} \cos{q k_x} \right],
\end{align}
where $p$ and $q$ are relative primes and $T_q(g)$ is the
Chebyshev polynomial of first kind of degree q. We can then simplify the left
hand side of Eq. \ref{eq:det5} and obtain the following,
\begin{align}
\prod_{\alpha=0}^{q-1} 2\left(M - (-1)^{\nu} t^{(1)}_z - t_x \cos{
\left(k_x + \frac{2\pi p}{q} \alpha\right)}\right) = 2 (-t_x)^q \left[
T_q(g) + (-1)^{p+q} \cos{q k_x} \right]
\end{align}
where we identify $g = \left(-M + \left(-1\right)^{\nu} t^{(1)}_z \right)/t_x$.
Rewriting Eq. \ref{eq:det5} with the above simplification, we get the following
expression for $k_x$
\begin{align}\label{eq:gapCondition1}
\cos{qk_x} = (-1)^{p+q}\left[ - T_q(g) + (-1)^{\mu -q}~ 2^{q-1} ~ (t_x/t_y)^{-q} \right].
\end{align}
Note that Eq. \ref{eq:gapCondition1} involves only the  two parameters 
$(g, t_x/t_y)$. For a given $(g, t_x/t_y)$, if $k_x=k_0$ is a solution of Eq.
\ref{eq:gapCondition1}, then $k_x=-k_0$ too satisfies the same equation. Further,
since $\cos{q(\pm k_0 + 2\pi m/q)} = \cos{qk_0}$, $k_x=k_0 + 2\pi m/q$,
$m$ runs over $0, 1, 2, ...,(q-1)$; and $k_x=-k_0 + 2\pi m/q$, where now $m$
runs over $ 1, 2,...., q$; are also solutions of Eq. \ref{eq:gapCondition1}.
We  recall that $k_x$ lives in the magnetic BZ, $0\leq k_x \leq 2\pi$,  and in the
above $k_0$ is restricted to $0\leq k_0 \leq 2\pi/q$. Hence, for a given $\mu$ and $\nu$,
there are a  total 
$2q$ number of $k_x$ values in the magnetic BZ where  the spectrum is gapless.
We have numerically verified that these  distinct gapless points in magnetic BZ
are the Weyl nodes in the theory. 

We can also obtain the boundaries of the topological phases from Eq.
\ref{eq:gapCondition1}. Note that the gapless point $k_0$ could be anywhere
in between 0 and $2\pi/q$. But $\cos{qk_0}$ is distinct only in the range
$0\leq k_0 \leq \pi/q$. So when we vary $k_0$ from 0 to $\pi/q$, the Eq.
\ref{eq:gapCondition1} describes gapless regions in the parameter space.
Solutions of Eq. \ref{eq:gapCondition1} exist only if the value of the 
RHS lies in the interval $[-1,1]$. For a given $\mu$ and $\nu$,
the spectrum has  $2q$ zeros (gapless points in BZ) when RHS $\in (-1,1)$.
As explained in the argument before Eq. \ref{eq:B0gapless}, phase
transitions, which imply a change in the number of solutions for a
particular $\mu,\nu$, can occur only when the RHS is at the edge of
its allowed range, namely $\pm1$. Therefore the  phase boundaries in
the  $(g, t_x/t_y)$ space are given by
\begin{subequations}
\label{eq:critical}
\begin{align}
\label{eq:critical1} - T_q(g) + (-1)^{\mu -q} ~ 2^{q-1}~ (t_x/t_y)^{-q} =& ~ (-1)^{q+p} \\ 
\label{eq:critical2} - T_q(g) + (-1)^{\mu -q} ~ 2^{q-1}~ (t_x/t_y)^{-q} =& -(-1)^{q+p}.
\end{align}
\end{subequations}
Eqs. \ref{eq:critical1} and \ref{eq:critical2} are obtained from Eq.
\ref{eq:gapCondition1} by setting  the RHS equal to $1$ and $-1 $
respectively.  Note the dependence of $p$ in the above equations; 
it only matters whether $p$ is even or odd. But even and odd
values of $p$ merely interchange the  equations Eq. \ref{eq:critical1} and Eq.
\ref{eq:critical2}. So $p$ does not affect the topological phase
diagram at all. Each of the above equations is a set of four equations,
because both $\mu$ and $\nu$ take values  0 and 1 ($\nu$ enters through $g=(-M +
(-1)^{\nu} t^{(1)}_z)/t_x$). Note  that the phase boundaries given in Eq.
\ref{eq:critical} are solely determined by only two parameters
$(g, t_x/t_y)$. Therefore, the  topological phase diagram is essentially
controlled by the  two parameters $(M/t_y, t_x/t_y)$. The role of the 
parameter $t^{(1)}_z$ which enters through $g$ is  to merely  shift the
origin of $M$  - it does not lead to any  new phase. The topological phase diagram is 
shown in Fig. \ref{fig:phase}. There are gapless WSM phases,  the W2$'$  phase, a
 gapped Layered Chern Insulator (LCI) phase, an unusual  $I'$ insulator phase 
 along with a trivial insulator phase.

\section{Critical $M^c$ and $t^c_x$  for  LCI phase}
\label{appendix:criticalLCI}

The critical  values  $M^c$ and  $t^c_x$ which are defined in the main 
text in the Sec. \ref{sec:LCIphase}  can be derived as follows.  From the
graphical visualization of the critical curves (phase boundaries) which
can be seen  explicitly in the figures in Fig.\ref{fig:phase} in the main text,
we find that $t^c_x$ and $M^c$  are  given by the intersection of the 
following two critical curves (obtained by putting $\mu=1$, $\nu=0$ in 
Eq. \ref{eq:critical1} and $\mu=0$, $\nu=0$ in Eq. \ref{eq:critical2}) 
\begin{subequations}
\begin{align}
\label{eq:critical3} - T_q(g) + (-1)^{1 -q} ~ 2^{q-1}~ (t_x/t_y)^{-q} = & ~ (-1)^{q+1} \\ 
\label{eq:critical4} - T_q(g) + (-1)^{-q} ~ 2^{q-1}~ (t_x/t_y)^{-q} = & ~ - (-1)^{q+1} , 
\end{align}
\end{subequations}
where, now, $g=(-M + (-1)^{\nu} t^{(1)}_z)/t_x =(-M + t^{(1)}_z)/t_x $
and we have used $p=1$. We can easily find $t^c_x$ by subtracting Eq.
\ref{eq:critical3} from Eq. \ref{eq:critical4} and  the  corresponding $M^c$
can be found by adding the two equations to get $T_q(g)=0$ and   solving  for
$M^c$. We find the following solutions for $q>1$,
\begin{align}
t^c_x &= t_y \hspace{0.1cm} 2^{1-1/q} \\
M_c &= t^{(1)}_z + t^c_x \cos{\left(\pi/2q\right)}= \left(1 + 2^{1-1/q} \cos{(\pi/2q)}\right) t_y. 
\end{align}
For $q=1$,  we have  $t_x^c  = t_y$ and $M^c = 2t_y$ (where we have used 
$t_z^{(1)} = t_y$). In the $q\to \infty$ limit, these critical values approach to 
$M^c = 3t_y$ and $t_x^c = 2t_y$. This set of values gives the region $M>M^c,
t_x >t_x^c$, where the LCI phase cannot appear.

\bibliography{referencesLchern}
\end{document}